\pdfoutput=1

\documentclass[10pt,journal,compsoc]{IEEEtran}

\usepackage{booktabs}
\usepackage{graphicx}
\usepackage{epstopdf}
\usepackage{subcaption}
\usepackage{multirow}
\usepackage{amsmath,amssymb}
\usepackage[lined,boxed,vlined,ruled,linesnumbered]{algorithm2e}
\usepackage{caption}
\usepackage{tikz}
\usepackage{pgflibraryarrows}
\usepackage{pgflibrarysnakes}
\usepackage[utf8]{inputenc}
\usepackage[english]{babel}
\usepackage{hyperref}
\usepackage{dsfont}
\usepackage{bbm}
\usepackage{url}
\usepackage{enumitem}
\usepackage{makecell}
\usepackage{mathrsfs}
\usepackage{booktabs}
\usepackage{url}
\usepackage{amsmath}
\usepackage{amssymb}%花体字母加粗
\usepackage{mathrsfs}%花体字母

\newtheorem{theorem}{Theorem}
\newtheorem{definition}{Definition}
\newtheorem{corollary}{Corollary}

\makeatletter
  \newcommand\figcaption{\def\@captype{figure}\caption}
  \newcommand\tabcaption{\def\@captype{table}\caption}
\makeatother

\newcommand{\stitle}[1]{\vspace{1ex}\noindent\textup{\textbf{#1.}}}

\begin{document}
\title{A General Early-Stopping Module for Crowdsourced Ranking}

\author{Caihua~Shan,
		Leong~Hou~U,
        Nikos~Mamoulis,
        Reynold~Cheng,
        and~Xiang~Li

\IEEEcompsocitemizethanks{\IEEEcompsocthanksitem C. Shan, R. Cheng and X. Li are with Department of Computer Science, University of
Hong Kong, Pokfulam Road, Hong Kong.
E-mail: {chshan, ckcheng, xli2}@cs.hku.hk

\IEEEcompsocthanksitem L.H. U is with State Key Laboratory of Internet of Things for Smart City,
Department of Computer and Information Science,
University of Macau, Macau.
E-mail: {ryanlhu}@umac.mo

\IEEEcompsocthanksitem N. Mamoulis is with the Department
of Computing Science, University of Ioannina, Ioannina, Epirus, Greece. 
E-mail: {nikos}@cs.uoi.gr
}
\thanks{Manuscript received xx xx. 20xx; revised xx xx. 20xx; accepted xx xx. 20xx}
}

%}%Manuscript received March 13, 2018; revised August 26, 2018.}}
\markboth{Journal of \LaTeX\ Class Files,~Vol.~14, No.~8, August~2019}%
{Shell \MakeLowercase{\textit{et al.}}: Bare Demo of IEEEtran.cls for Computer Society Journals}
\IEEEtitleabstractindextext{%
%!TEX root = main.tex

\begin{abstract}
Crowdsourcing can be used to determine a total order for an object set (e.g., the top-10 NBA players) based on crowd opinions. This ranking problem is often decomposed into a set of microtasks (e.g., pairwise comparisons). These microtasks are passed to a large number of workers and their answers are aggregated to infer the ranking. The number of microtasks depends on the budget allocated for the problem. Intuitively, the higher the number of microtask answers, the more accurate the ranking becomes. However, it is often hard to decide the budget required for an accurate ranking. We study how a ranking process can be terminated early, and yet achieve a high-quality ranking and great savings in the budget. We use statistical tools to estimate the quality of the ranking result at any stage of the crowdsourcing process, and terminate the process as soon as the desired quality is achieved. Our proposed early-stopping module can be seamlessly integrated with most existing inference algorithms and task assignment methods. We conduct extensive experiments and show that our early-stopping module is better than other existing general stopping criteria. We also implement a prototype system to demonstrate the usability and effectiveness of our approach in practice.
\end{abstract}

\begin{IEEEkeywords}
Crowdsourcing, Ranking, Early Stopping.
\end{IEEEkeywords}}

\maketitle

\IEEEdisplaynontitleabstractindextext
\IEEEpeerreviewmaketitle
%!TEX root = main.tex

\section{Introduction}\label{sec:introduction}
\setlist{leftmargin=3mm}

Crowdsourcing has been used to address a variety of problems, such as language translation~\cite{translation}, entity matching~\cite{entityresolution,entityresolution2}, image labeling~\cite{labeling,videolabeling}, and object ranking~\cite{local,kou2017crowdsourced}.
These problems, which are typically hard for computers to solve, can be easier for humans. Crowdsourcing leverages both human and machine intelligence to derive a solution (e.g., correct inaccuracies in machine predictions).  In this paper, we study the use of crowdsourcing on {\em ranking} objects. This approach, which has received a lot of attention from different research communities~\cite{crowdgauss,ssrw,kou2017crowdsourced},
is particularly helpful when ranking cannot be done objectively.
For example, to determine the greatest athletes of all times or the best pictures of a landmark,
%(e.g., finding the greatest athletes of all times) should ask
we could solicit opinions from the crowd
% to compare between great athletes
and aggregate them to a ranking that maximizes the consensus.
%comparisons to
%the opinion a broad range of crowd to increase the level of consensus and representative.
In addition, crowdsourced ranking can be used to filter data for subsequent machine learning tasks. For instance, ranking answers to a question posted in a forum
% to a Stack Overflow question
and selecting only the top ones can ease the burden of natural language processing.
%and recommendation analysis.
%\nikos{this example may not be good because Stack Overflow answers already get votes from the crowd which is much easier than the comparison setup that we have; I think the reviewer can easily say that this problem is solved already.}
%\caihua{ Other application? In image search, given 1000 photos of a restaurant, find top-$k$ list photo that are most appealing and best describe the restaurant \cite{local}
%}

To conduct crowdsourced ranking, existing solutions typically decompose the ranking process into a set of small and easy-to-answer microtasks, such as pairwise comparisons~\cite{exptopk}.
%or rating question.%Likert scale rating.
%\nikos{cite some papers here in order to convince the reviewer that this is what is done typically} \caihua{
The microtasks are then distributed via crowdsourcing platforms, such as Amazon Mechanical Turk (AMT)~\cite{AMT} and FigureEight~\cite{figureEight}, to crowd workers by offering incentives, e.g., money, reputation, etc. The final ranking is computed by an inference algorithm based on the answers collected from the crowd. Naturally, the ranking accuracy is proportional to the number of collected answers to microtasks, i.e., the {\em total budget} paid by the requester.

\begin{figure}[t!]
  \centering
  \includegraphics[width=1\columnwidth]{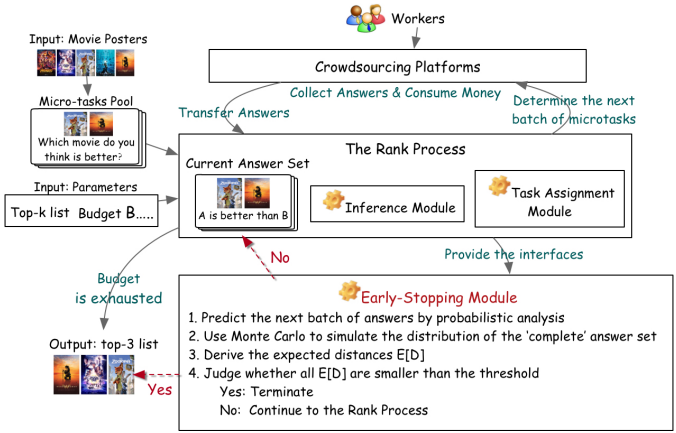}
  \caption{Crowdsourced ranking framework}
  \label{fig:general_process}
\end{figure}

Recent studies~\cite{crowdbt,local,ssrw} attempt to improve the {\em inference algorithm} $\mathcal{I}$ and fine-tune the {\em task assignment} $\mathcal{T}$ (i.e., by dispatching tasks to suitable workers), in order to spend the budget more effectively.
Typically, the microtask answers are collected in batches. Let $A_i$ be the $i$th batch of answers;
Inference algorithm module $\mathcal{I}$ infers the {\em interim ranking} from $A_1\cup ... \cup A_i$;
Task assignment module $\mathcal{T}$ is used to determine the next batch of microtasks and assign them to crowdsourcing platforms.
Fig.~\ref{fig:general_process}
illustrates this {\em ranking process} $\mathcal{R}$.
%For the ease of understanding, we illustrate the processing framework in Fig.~\ref{fig:general_process}.

According to a recent experimental survey on crowdsourced ranking~\cite{exptopk}, there is no single winner method that outperforms all others in all performance factors
(accuracy, convergence rate, efficiency, scalability). In addition, most approaches require the budget to be set in advance, but they offer no guideline on how to set this value. Hence, it is expected that the requester sets a large enough budget, hoping that the ranking process will converge to a stable ranking. This raises an interesting question: {\em can we spend less and achieve approximately the same ranking, as if we had spent all the budget?}

\begin{figure}[t!]
  \centering
  \includegraphics[width=0.95\columnwidth]{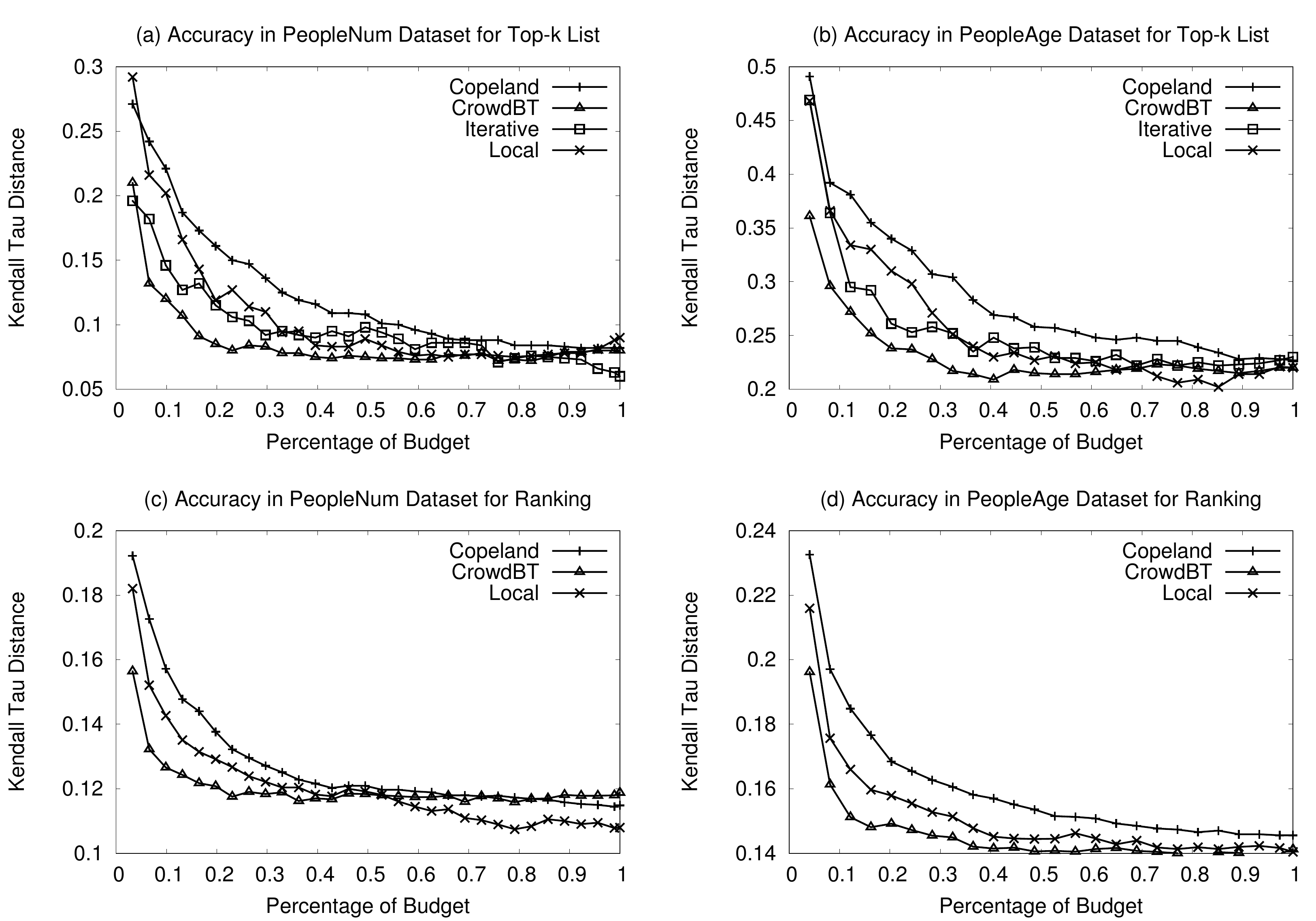}
  \caption{Accuracy versus Budget}
  \label{fig:observation}
\end{figure}

To answer this question, we first investigate how much budget could be saved when some representative inference algorithms are applied, i.e., Copeland~\cite{Copeland}, CrowdBT~\cite{crowdbt}, Iterative~\cite{local}, and Local~\cite{local}.
Details about these methods are given in Sec.~\ref{sec:experimental_setting}.
% The introduction of these methods can be found in the related work section (Sec.~\ref{sec:relatedwork}).
%Fig.~\ref{fig:observation} shows how the {\em accuracy} of these algorithms changes as a percentage of the budget is used up by varying the consumed budget.
We carry out the ranking and top-$10$ query tasks on two public datasets, namely peopleAge~\cite{exptopk} and peopleNum~\cite{datasetPeopleNum}.
Fig.~\ref{fig:observation} shows how the {\em accuracy} of these algorithms varies as the budget increases. 
%We use the number of answers in the x-axis to represent how much budget consumed. 
As an accuracy measure, we utilize Kendall's tau distance between the rankings progressively inferred and the ground truth ranking.
%As shown in Fig.~\ref{fig:observation}, the accuracy of all methods keep improving when we increase the budget. Besides,
All methods converge to a {\em stable state}\footnote{A formal definition of the stable state is provided in the Sec.~\ref{sec:stopping_point}.}, where the change of the distance induced by the inferred ranking is very small. 
%becomes close to that of the final ranking, at the point where all the budget is used up. 
%that of the final state, before running out of our budget.
In Fig.~\ref{fig:observation}(a) and (b), CrowdBT reaches a stable state after using just 40\% of the budget, whereas all other methods converge when 60\% of the budget is used. 
%In Fig.~\ref{fig:observation}(a) and (b), CrowdBT reaches a stable state after using just 40\% of the budget ($\Delta$distance is no more than 0.01), whereas all other methods converge when 60\% of the budget is used ($\Delta$distance is no more than 0.03).
Similarly for ranking in Fig.~\ref{fig:observation}(c) and (d), the distance changes are all smaller than 0.02 after using 50\% of the budget. 
Obviously, we can {\em stop early} the crowdsourcing process when we
%can correctly predict the point where we
reach a  stable state.
We now face the following challenge: {\em how do we know whether the ranking process has reached a stable state?}

%\caihua{Generally speaking, the stable state is that the change of the distance (y-axis in Fig.~\ref{fig:observation}) during a certain range of the consumed budget (x-axis in Fig.~\ref{fig:observation}) is smaller than a threshold. To guarantee the accuracy of the final result, we also require this stable state must maintain until the final state when all the budget exhausts. However it is not easy to find such state because we only know the collected answers but there exists the randomness which comparisons will be asked and their answers in the future. Thus we do not know the final state exactly. Besides, the distance between interim rankings also fluctuates, caused by the noise from answers and the order of comparisons between objects. It increases the difficulty of detecting this state.}

%\nikos{the paragraph above introduces some confusion, so I suggest to remove it. I have added a footnote explaining that the stable state will be defined in section 2 in order to comfort the reader.}

To tackle this challenge, we develop a novel Early-Stopping (ES) module that attempts to predict the next batch of answers by probabilistic analysis.
We then use Monte Carlo simulation~\cite{MonteCarlo}, based on the prediction model, to construct the distribution of the final answer
and, in turn, derive the {\em expected accuracy} of the final state. This helps us to assess when the ranking process reaches its stable state, subject to a budget $B$. To early-stop the process, our ES module requires an {\em accuracy tolerance} $\theta$ parameter, i.e., the acceptable accuracy that we can afford to lose when compared to the ranking that will be obtained if all the budget is used up.

Our ES module can seamlessly be used by most ranking processes with minimal effort. The only requirement is that the process provides interfaces for the inference and task assignment modules,
%the algorithm shares its inferred interim ranking results (using the collected answers so far $\mathcal{L}_1\cup...\cup\mathcal{L}_i$)
and accepts a programming call to terminate the crowdsourcing process, when our module determines that the expected accuracy already satisfies tolerance $\theta$. We emphasize that the development of our ES module is orthogonal to that of inference algorithms and task assignment methods.
%This means that our module can be easily added into a system that supports crowdsourced ranking queries.

The main contributions of this paper are summarized as follows:

%Different from them, our stopping criterion is general and users could set the meaningful parameters to control the stopping time. We use some statistic properties to handle the uncertainty for worker answers. Our module could embed into the existing collection process in crowdsourced platforms. We treat aggregation and assignment method as a black box which users could choose an appropriate method by themselves. The things we need from users are their maximum affordable budget and acceptable error. We predict the final result using the maximum affordable budget and stop when the difference between the current and the final result is smaller than maximum acceptable error. The main contribution of this paper can be summarized as follows:

\begin{itemize}

  \item To the best of our knowledge, we are the first to propose a general Early-Stopping (ES) module for crowdsourced ranking.
  \item Our ES module is orthogonal to any inference algorithm or task assignment method, and does not interfere with the flow of the crowdsourced ranking process.
%    ; hence, its development is orthogonal to that of the inference algorithms. In short, our module can be viewed as a module built on top of any inference algorithms.
  \item We thoroughly evaluate our ES module with subjective and objective tasks, different inference algorithms and task assignment methods, varying budgets and accuracy tolerances. Our module can save even half of the budget given to the ranking processes.
  \item We have developed a prototype system and conducted an online experiment in AMT to assess the effectiveness of our ES module.
%    Given the budget $B$ and the accuracy tolerance $T_{acc}$, our ES module is able to recommend an early stopping point of the inference algorithm subject to $T_{acc}$. According to our experiments, the requestor could save more than half of their initial budget.

\end{itemize}

The rest of the paper is organized as follows. We formulate the problem and provide definitions and notations in Sec.~2. Our ES module is described in detail in Sec.~3. The experimental evaluations are shown in Sec.~4 and online experiments using our system prototype is presented in Sec.~5. We discuss related work in Sec.~6 and conclude in Sec.~7.

% \begin{figure}[t!]
%   \centering
%   \begin{subfigure}[b]{0.5\columnwidth}
%     \includegraphics[width=\textwidth]{figs/observation_a.eps}
%     \caption{PeopleNum ($k=10$)}
%   \end{subfigure}%
%   \begin{subfigure}[b]{0.5\columnwidth}
%     \includegraphics[width=\textwidth]{figs/observation_b.eps}
%     \caption{PeopleAge ($k=10$)}
%   \end{subfigure}
% \caption{Accuracy versus Budget}
% \label{fig:observation}
% \end{figure}

\section{Preliminaries}\label{sec:preliminary}
\setlist{leftmargin=10mm}

%In this section, we define the crowdsourced rank query and the crowdsourced  top-$k$ query. We will also introduce the detail of crowd data collection framework.

%and describe a detailed general collection process to obtain them.
%We also define the uncertainty of current answers.

%\subsection{Definitions} \label{sec:definition_rank_topk_list}

We first define crowdsourced ranking and top-$k$ queries as follows.

\begin{definition}[\textbf{Crowdsourced Ranking}]
Given a set of $n$ objects $\mathcal{O}=\{o_1,...,o_n\}$, use human workers to decide a total order $\sigma=\{o_i\prec o_j \prec ...\}$.
\end{definition}

%which are hard to compare by a computer,
\begin{definition}[\textbf{Crowdsourced Top-$k$ Query}]
  Given a set of $n$ objects $\mathcal{O}=\{o_1,...,o_n\}$, use human workers to find a ranked list $\sigma^k=\{o_i\prec o_j \prec ...\}$ of size $k$, such that for any $o_i\in \sigma^k \land o_l \notin \sigma^k$, $o_i \prec o_l$.
\end{definition}

Note that the operator $\prec$ is a conclusion drawn from the crowd's answers. For instance, given some replies to a question posted in a forum, we can ask
the crowd to conduct pairwise comparisons between the replies, and then
use existing inference algorithms to process the  crowd's input and find the top-5 replies. Note that comparing two replies is not machine friendly since it not only requires strong natural language processing techniques but also a good understanding of the question, i.e., domain expertise.

%\begin{definition} (\textbf{Multi-wise Comparison}).
% The worker $w_{i}$ is required to choose the best object given a set of $m$ objects $\mathcal{S}_i=({o}_{i1},{o}_{i2},...,{o}_{im})$. The answer $a_{i}$ is the index of her most preferred object, which could be $m$ cases, 0, 1,..., m-1. Current set of answers from multi-wise comparison can be also represented as a set of $3$-tuples $\mathcal{L}_c={\{(\mathcal{S}_i,w_{i},a_{i})\}}_{ i=1...|\mathcal{L}_c|}$.
% \end{definition}

\subsection{Distance Between Rankings} \label{sec:distance}

In our solution, we need to measure the distance (i.e., difference) between the ranking inferred at an intermediate state and the ranked list at the final state. To measure the distance between two rankings, a common practice is to use {\em Kendall's tau distance}, i.e., the number of inverse pairs of objects.% in one ranking required to make it identical to the other ranking.
%\nikos{confirm that this is correct and if yes delete this comment. Your description was not accurate} \caihua{Kendall’s tau is equal to the number of exchanges needed in a bubble sort to convert one ranking to the other.}

We use the normalized Kendall's tau distance for complete rankings and top-$k$ ranked lists as defined in Eq.~\ref{eq:ranklist} and Eq.~\ref{eq:topklist}, respectively:

\begin{equation}\label{eq:ranklist} \small
    \mathbb{D}(\sigma_1, \sigma_2) = \frac{\sum_{(o_i,o_j)\in O \times O, i<j} \mathds{1}(o_i \prec o_j, \sigma_1) \times \mathds{1}(o_i \succ o_j, \sigma_2) }{n(n-1)/2}
\end{equation}
\begin{equation}\label{eq:topklist}\small
    \mathbb{D}(\sigma^k_1, \sigma^k_2) = \frac{\sum_{(o_i,o_j)\in O \times O, i<j} \mathds{1}(o_i \prec o_j, \sigma^k_1) \times \mathds{1}(o_i \succ o_j, \sigma^k_2) }{k^2}
\end{equation}
where $\mathds{1}$ is the indicator function that equals to 1 when its predicate is true, or 0 otherwise.
When $\sigma_1$ and $\sigma_2$ are reversed, the numerator of Eq.~\ref{eq:ranklist} takes its maximum possible value $n(n-1)/2$, and Eq.~\ref{eq:ranklist} reaches the highest value of 1.
As for Eq.~\ref{eq:topklist}, the numerator takes its maximum value $k^2$ when objects in $\sigma^k_1$ and $\sigma^k_2$ have no intersection.
%of the numerator in $\mathbb{D}(\sigma^k_1, \sigma^k_2)$ is $k^2$
%\nikos{are you sure the maximum is $k^2$? I though the maximum is 1. $k^2$ means that the numerator is $k^4$.}
%\caihua{The maximum value of the numerator in  $\mathbb{D}(\sigma^k_1, \sigma^k_2)$  is $k^2$.}

Note that there is a case when both $o_i$ and $o_j$ are in one ranked list (e.g., $\sigma^k_1$), and none of them is in the other list (e.g., $\sigma^k_2$). In this case, we have to guess the position of $o_i$ and $o_j$ in $\sigma^k_2$. We select the optimistic attitude, i.e., $o_i$ and $o_j$ are in the same order in both top-$k$ lists, and do not impose any penalty in the distance function.

\begin{table}[t!]
\centering
\caption{Table of Notations}    
\label{table:notation}
\begin{tabular}{|@{~}c@{~}|@{~}c@{~}|}
\hline
\textbf{Notation} & \textbf{Description} \\ \hline \hline
$\mathcal{I}$         & Inference module     \\ \hline
$\mathcal{T}$         & Task assignment module     \\ \hline
$\mathcal{R}$         & Rank process containing $\mathcal{I}$ and $\mathcal{T}$     \\ \hline
$\mathbb{D}$          & Distance function for rankings     \\ \hline
$B$                   & Budget    \\ \hline
$\theta$              & Accuracy tolerance     \\ \hline
$n_\text{batch}$        & Number of tasks in a batch     \\ \hline
$A_1, ... ,A_i$   &  Answers obtained in the $i$th batch     \\ \hline
$\sigma_1, ... ,\sigma_i$   & Rank list calculated by  $\mathcal{I}(A_1 \cup ...\cup A_i)$    \\ \hline
$p_1, ... ,p_i$   &  State after collecting $i$ batches of answers     \\ \hline
$A^c / \sigma^c$      & Current answer set / Current rank list  \\ \hline
$\mathrm{E}[\mathbb{D}_{ij}]$ & Expected distance between the $i$th \& $j$th batches \\ \hline
$\overline{\mathbb{D}}_{ij}$ &  Mean of sampled distances between the $i$th \& $j$th batches \\ \hline
%$\mathbb{A}=\{A^1, A^2, ...\}$  & Possible worlds of the complete answer sets     \\ \hline
%$\Sigma=\{\sigma^1,\sigma^2, ...\}$ & Possible worlds of the final rank lists     \\ \hline

\end{tabular}
%\vspace{-1em}
\end{table}

% \begin{figure}[t!]
%   \centering
%   \includegraphics[width=0.7\columnwidth]{figs/fake_observation.pdf}
% \caption{Example of Stable State \& Optimal Stopping Point}
% \label{fig:fake_observation}
% \end{figure}

\subsection{Stable State \& Optimal Stopping Point} \label{sec:stopping_point}
Publishing a batch of microtasks into the crowdsourced platform is a common strategy to accelerate the speed of collections. Let $p_i$ be the state after collecting the $i$th batch of answers $A_i$ and $\sigma_{i} = \mathcal{I}(A_1 \cup ... \cup A_i)$ is the ranked list at $p_i$.
The stopping module of the crowdsourced ranking algorithm should check whether to stop at each $p_i$.
Without loss of generality, we assume that the budget $B$ is the total number of microtasks we plan to publish and the number of microtasks, $n_\text{batch}$, is the same in each batch.
%The budget $B$ can be divided evenly by $n_\text{batch}$ and $B/{n_\text{batch}}$ is the total number of batches needed to collect all answers.
$B/{n_\text{batch}}$ is the total number of batches needed to collect all answers.
We then give a formal definition of the stable state that we mentioned in the Introduction:

\begin{definition}[\textbf{Stable State}] \label{def:stable_point}
Given the whole collection process $\{A_1, A_2, \cdots, A_{B/n_\text{batch}}\}$ and an accuracy tolerance $\theta \in [0,1]$ from the requester, $p_l$ is called as a {\em stable state} of the process if:
%\begin{itemize}
\begin{enumerate}
  \item $\forall p_i, p_j \in [p_l, p_\text{final}],~\mathbb{D}(\sigma_i,\sigma_j) \leq \theta$
  \item $\nexists ~p_i < p_l,~p_i$ is a stable state
  %\item $\nexists \text{ a state } [p_{l'}, p_{r'}] \text{ is the stable state and }  [p_l, p_r] \subset [p_{l'}, p_{r'}]$
\end{enumerate}
where $l \in [1, B/n_\text{batch}]$ and $p_\text{final}=B/n_\text{batch}$.
%\end{itemize}
%Moreover, there does not exist another stable state $[p_{l'}, p_{r'}]$, which satisfies $[p_l, p_r] \subset [p_{l'}, p_{r'}]$.
\end{definition}

The first condition secures that the distances between the rankings at any two states (from $p_l$ to the final) do not exceed $\theta$. The second condition secures the maximality that no earlier (better) stable state can be found in the entire process. It is obvious only one stable state exists in each collection process.

%The stopping point $p_{\text{sc}}$ is the moment decided by a stopping criterion (SC) to early stop the ranking process. Based on the stable interval definition, we can define the optimal stopping point as follows:
%\begin{corollary}[\textbf{The Optimal Stopping Point}] \label{def:best_stopping_point}

The stopping point $p_{\text{sc}}$ is the moment decided by a stopping criterion (SC) to early stop the ranking process. Based on the stable state definition, we can say that 
\begin{corollary}[\textbf{The Optimal Stopping Point}] \label{def:best_stopping_point}
The optimal point $p_{\text{optimal}}$ to early stop a ranking process is when the process turns into the stable state, i.e., $p_{\text{optimal}}=p_l$.
\end{corollary}

%Hence, we say a process at state $p_l$ turns into {\em stable} if the distances between the rankings at any two subsequent states (i.e., $[p_l,p_\text{finsh}]$ do not exceed $\theta$.
%the stable interval is an execution interval until the final execution state, such that the distances between the rankings at any two states do not exceed $\theta$.
%Besides, the definition secures that the stable interval satisfies the maximality, where there is no larger stable interval containing the  stable interval.
%It is obvious only one $p_l$ exists in each collection process.

%The stopping point $p_{\text{sc}}$ is the moment decided by a stopping criterion (SC) to early stop the ranking process. Based on the stable interval definition, we can define the optimal stopping point as follows:
%\begin{definition}[\textbf{The Optimal Stopping Point}] \label{def:best_stopping_point}
%The optimal stopping point $p_{\text{optimal}}$ is the starting point $p_l$ of the stable interval.
%\end{definition}

The optimal stopping point guarantees the optimality because it saves up as much as possible the budget and ensures the distances from the ranked list at the stopping point to the final ranking are always smaller than the accuracy tolerance $\theta$.

\begin{figure}[t!]
  \centering
  \begin{subfigure}[b]{0.485\columnwidth}
    \includegraphics[width=\textwidth]{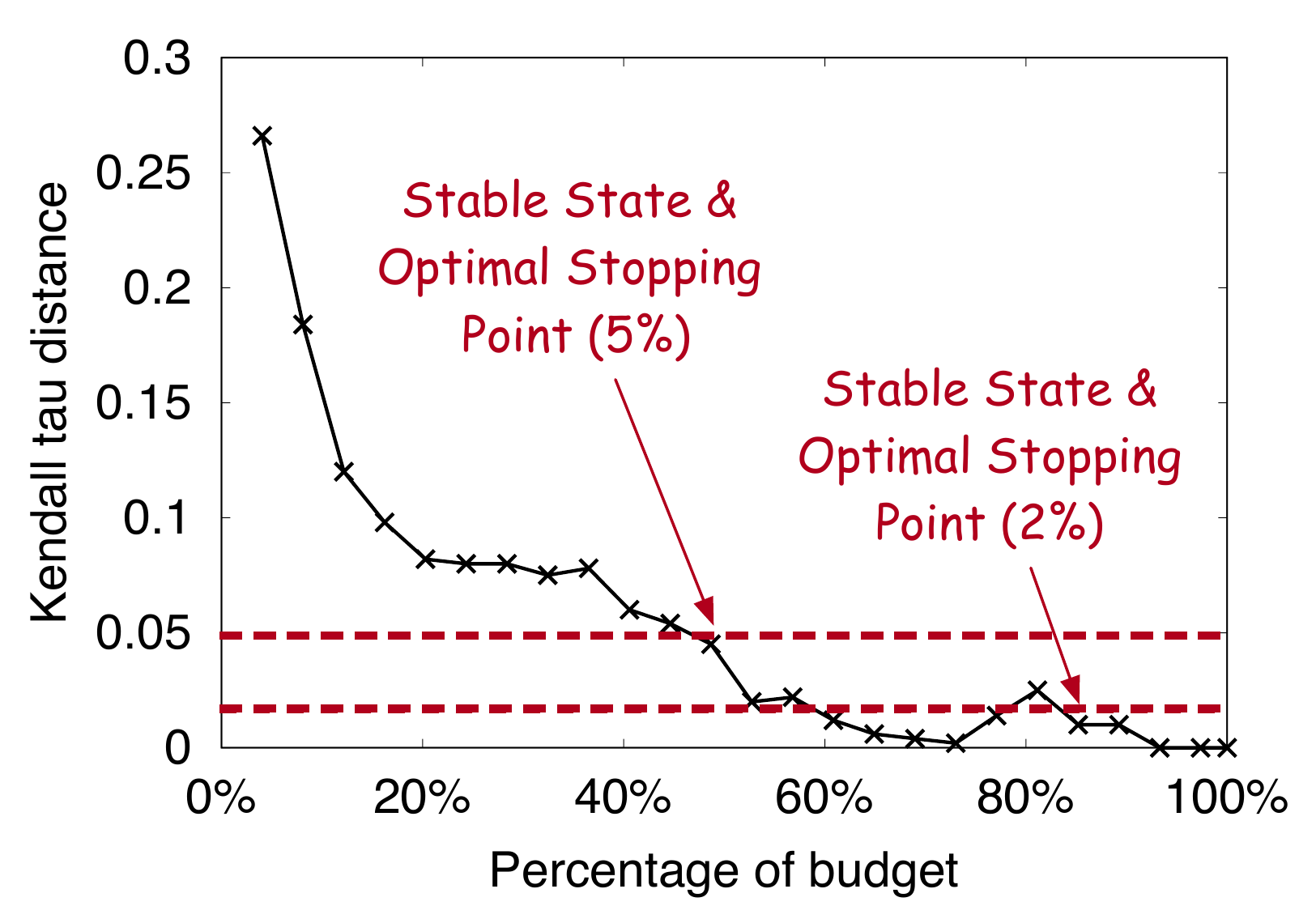}
    \caption{}
  \end{subfigure}%
  \begin{subfigure}[b]{0.485\columnwidth}
    \includegraphics[width=\textwidth]{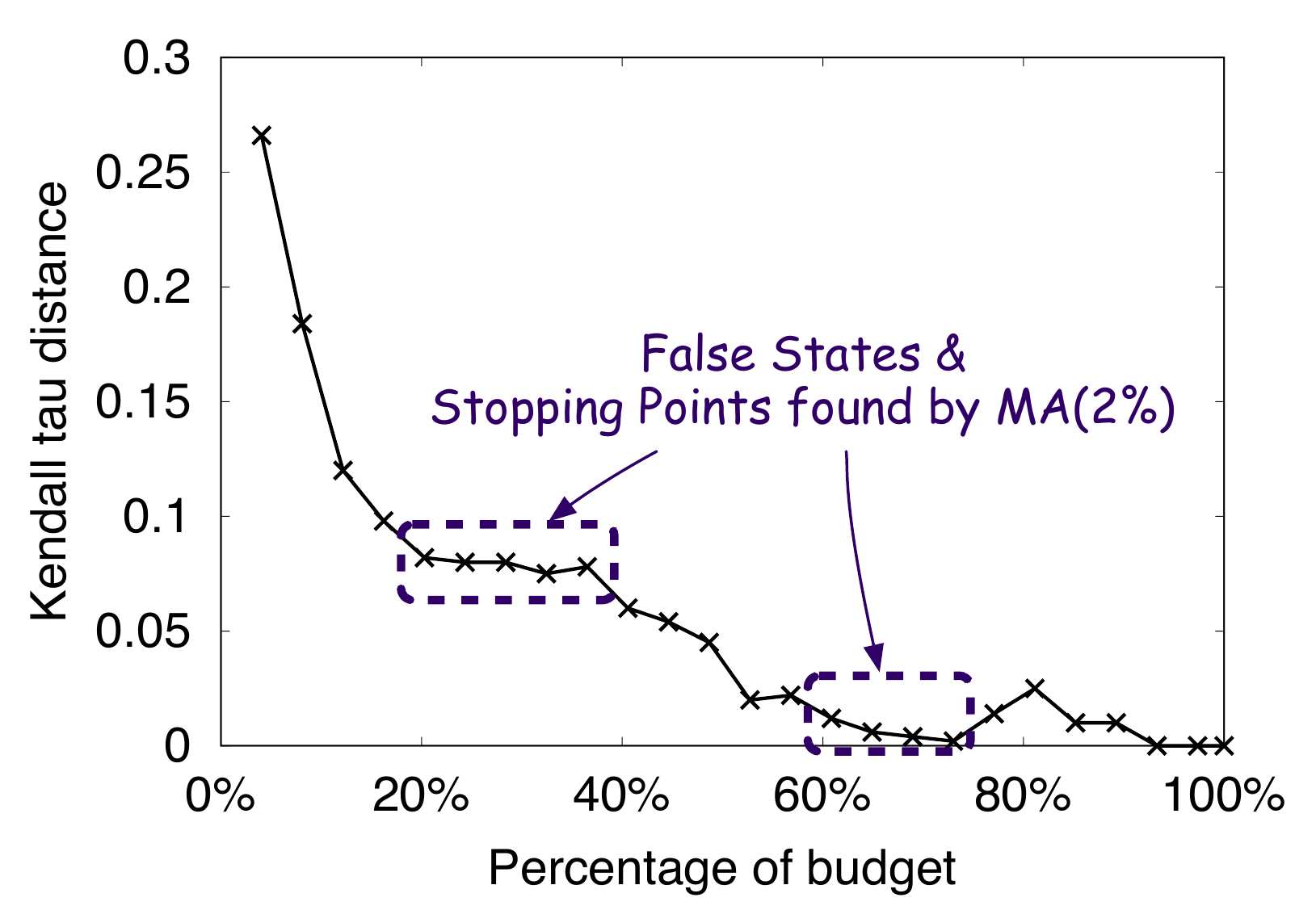}
    \caption{}
  \end{subfigure}
\caption{Examples of Stable State \& Optimal Stopping Point}
\label{fig:fake_observation}
\end{figure}

For example, Fig.~\ref{fig:fake_observation}(a) shows the distance between the current and the final ranking at all states of the process. We show two optimal stopping points with $\theta=5\%$ or $\theta = 2\%$. %Their corresponding stable  are from these points to the final execution state. 
Basically we may save more budget with larger $\theta$. Here we save $50\%$ budget for $\theta = 5\%$, and $10\%$ budget for $\theta = 2\%$.

%\nikos{does the stable state also have to satisfy maximality? Because for any stable state any subset of the interval is trivially a state. Also is there any minimum time length requirement for a stable state. Otherwise, two consecutive moments for which the distance is at most $\theta$ is trivially a state. Finally, it seems that the true stable state should definitely contain the last moment of the budget. This is not given as a requirement in the definition.}\caihua{add the maximality. }

%Different moments to stop are the trade-off between the accuracy of the result and the used budget.
%We care more about the accuracy because it is meaningless if spending the less money but obtaining the bad result. Thus, we define the oracle stopping point as follow:

%We use $p_\text{final} = B/n_\text{batch}$ as the final point. Compared with stable states, the end point of $p_{\text{optimal}}$ must be the final point. Besides $p_{\text{optimal}}$ should be the earliest moment satisfying the condition.

%It converges to a stable state with more budget.

One may wonder whether some simple method, e.g., Moving Average \cite{MovingAverage}, can find $p_{\text{optimal}}$. We show two kinds of intervals in Fig.~\ref{fig:fake_observation}(b), which has the same distance curve as Fig.~\ref{fig:fake_observation}(a). The first purple rectangle is an interval that tends to be stable during a certain time but descends gradually as more budget consumes. The second one also tends to be stable but the change of rankings is larger than $\theta$ as more budget consumes. Given a current point $p_i$, moving average uses the previous rank lists in a certain window size to represent the inferred rankings in the future. It is easy to drop out into these intervals and cause the process to stop earlier than it should. To avoid stopping at these intervals, we propose a novel ES module that attempts to discover the optimal stopping point. Table~\ref{table:notation} shows the notations used in this paper.

\section{Early-Stopping Module}\label{sec:approach}

%Our Early-Stopping (ES) module needs to estimate how far the current ranked list is from the final state. Given the budget $B$ and an accuracy tolerance $\theta$, the goal is to terminate collecting answers as soon as the current ranking satisfies $\theta$. 
We first discuss how to predict the next batch of answer set $A_{i+1}$ by probabilistic analysis. To estimate the final state of the ranking, we use the Monte Carlo method to generate the different possible worlds of the complete answer sets, estimate the expected distances and judge whether stop or not. 
%$\mathbb{A}=\{A^1, A^2, ...\}$. Based on them, we estimate the expected distance from the interim ranked list to possible final rankings. 

\subsection{Predicting the Next Answer Set} \label{sec:predict_remain_ans}

Consider a crowdsourcing rank process $\mathcal{R}$, based on an inference module $\mathcal{I}$ and a task assignment module $\mathcal{T}$, that has already collected the $i$th batch answer set ($A^c=A_1\cup...\cup A_i$).
We predict the next batch of answers by a three-stage process, including (1) determining new tasks ${t_{new}}$, (2) predicting the answers $a_{new}$ of $t_{new}$, and (3) estimating the influence of worker reliability to the answers $a_{new}$.

%Given the number of answers $r$ we want to predict, there are three challenges to predict the next answer set $\mathcal{R}$: 1) which candidate micro-tasks will be selected, 2) which worker will come and 3) what the answer is. Besides, we may collect answers in many batches if $r$ is large. We also need to consider the uncertainty of the collection process, e.g., the order-dependent relation of collected answers. 

\subsubsection{{\bf Determining new tasks, $t_{new}$}} 

Recall that the microtasks of crowdsourced ranking are pairwise comparisons $(o_i,o_j)$. Given the collected answer set $A^c$, 
%the inference module $\mathcal{I}$ will infer the intermedia rank list $\sigma$ and 
the task assignment module $\mathcal{T}$ decides the importance of tasks. The most important $n_\text{batch}$ tasks are distributed to crowdsourcing platforms as the next batch. 
We predict answers for these tasks in our subsequent prediction model.
%The same $n_\text{batch}$ tasks $\{t_{new}=(o_i,o_j)\}$ are chosen by our prediction model. 

\subsubsection{{\bf Predicting the answer, $a_{new}$}}\label{sec:predictanswer}

Given the collected answer set $A^c$ and a chosen task $t_{new}=(o_i,o_j)$, we want to predict the answer to $t_{new}$. 
%Note that the answer prediction is independent of the selected inference module $\mathcal{I}$ since the answer will be given by the crowd.
%In the following, we first discuss a prediction model based on {\em passive} task assignment, which simply publishes the tasks on the platforms and the tasks can be done by any workers. 
We assume that the workers are reliable since they have to obey the crowdsourcing platform policy, e.g., gain reputation via user feedback. Thereby, we can regard the answer $a_{new}$ of the task $t_{new}=(o_i,o_j)$ as a Bernoulli distribution of the probability of $o_i\prec o_j$, denoted as $P_{ij}$. Formally: 
\begin{equation}\label{eq:anew}
a_{new} \sim \text{Bernoulli}(P_{ij}),
\end{equation}
where $P_{ij}$ is the probability of $o_i \prec o_j$.
Several models for $P_{ij}$ has been suggested in previous crowdsourcing studies~\cite{mallowsModel,BTL,thurstoneModel}.
For instance, the Bradley-Terry (BT) model~\cite{BTL} defines $P_{ij} = \frac{e^{s_i}}{e^{s_i}+e^{s_j}}$, where $s_i$ is the latent score of object $o_i$. The Thurstonian model~\cite{thurstoneModel} defines $P_{ij} = \Phi({s_i}-{s_j})$, where $\Phi$ is the normal cumulative distribution function. However, some inference modules~\cite{eriksson2013learning, davidson2013using} do not build on the latent scores of objects. 

We attempt to design a new estimation model that is suitable for most inference modules. We estimate the probability $P_{ij}$ independently, i.e., $P_{ij}$ only based on the previous answer set of the task $(o_i,o_j)$. 
%by a maximum likelihood estimation process. 
Suppose that the current answer set is $A^c$; we build an observed matrix $M$, where $M_{ij}$ is the number of answers reporting $o_i \prec o_j$ in $A^c$. $P_{ij}$ depends on $M_{ij}$ and $M_{ji}$.

We use maximum a posteriori probability (MAP) to calculate $\hat{P}_{ij}$:
\begin{equation} \small
\begin{aligned}
\hat{P}_{\mathrm{MAP}}(M) &=\underset{P}{\operatorname{arg\,max}} \ Pr(P \mid M) \\
&=\underset{P}{\operatorname{arg\,max}} \prod_{i,j|i<j} \ Pr(P_{ij} \mid M_{ij}, M_{ji}) \\
&=\underset{P}{\operatorname{arg\,max}} \prod_{i,j|i<j} \ \frac{Pr(M_{ij}, M_{ji} \mid P_{ij}) \, Pr(P_{ij})}
  {\displaystyle\int_{0}^{1} Pr(M_{ij}, M_{ji} \mid p_{ij}) \, Pr(p_{ij}) \, dp_{ij}} \\
&\varpropto \underset{P}{\operatorname{arg\,max}} \prod_{i,j|i<j} \ Pr(M_{ij}, M_{ji} \mid P_{ij}) \, Pr(P_{ij})
\end{aligned}
\end{equation}

If we assume the prior distribution of $P_{ij}$ as $Beta(1,1)$ which is the conjugate prior for the Bernoulli distribution, the posterior distribution of $P_{ij}$ is 
%Since $a_{new}$ follows the Bernoulli distribution, the conjugate prior distribution of $P_{ij}$ can be written as a Beta distribution. Accordingly, we can say 
\begin{equation}
Pr(P_{ij} \mid M_{ij}, M_{ji}) \sim Beta(M_{ij}+1, M_{ji}+1)
\end{equation}
The reason behind using $Beta(1,1)$ is that we believe that we have equal probability to get either $o_i\prec o_j$ or $o_i \succ o_j$. It could also be interpreted as Laplace smoothing to avoid some undefined calculation, e.g., $Beta(0,0)$.
%where $M_{ij}$ and $M_{ji}$ are the numbers of ($a_{new}=1$) or ($a_{new}=0$), respectively, from the observed answers. Note that we add a constant value (i.e., +1) as a Laplace smoothing for the subsequent probabilistic estimation (avoiding some undefined calculation).

The MAP of $\hat{P}_{ij}$ equals the mode of the posterior distribution, which is
\begin{equation} \label{eq:pij}
\hat{P}_{ij} = \frac{M_{ij}+1}{M_{ij}+M_{ji}+2}
\end{equation}

Alternatively, we could also use maximum likelihood estimate (MLE) to calculate $\hat{P}_{ij}$:
\begin{equation} %\small
\begin{aligned}
\hat{P}_{\mathrm{MLE}}(M) &=\underset{P}{\operatorname{arg\,max}} \ Pr(M \mid P) \\
&=\underset{P}{\operatorname{arg\,max}} \prod_{i,j|i<j} \ Pr(M_{ij}, M_{ji} \mid P_{ij}) \\
%&=\prod_{i,j|i<j} \ \binom{M_{ij}+M_{ji}}{M_{ij}} \,{P_{ij}}^{M_{ij}}\, {(1-P_{ij})}^{M_{ji}}
&=\underset{P}{\operatorname{arg\,max}} \prod_{i,j|i<j} \ {P_{ij}}^{M_{ij}}\, {(1-P_{ij})}^{M_{ji}}
\end{aligned}
\end{equation}
The MLE of $\hat{P}_{ij}$ equals to $\frac{M_{ij}}{M_{ij}+M_{ji}}$. Similarly, if we replace $M_{ij}$ and $M_{ji}$ by $M_{ij}+1$ and $M_{ji}+1$, respectively, by the Laplace smoothing, then the MLE equation will be identical to Eq.~\ref{eq:pij} (from MAP).

In summary, we estimate $P_{ij}$ from $M$ based on $A^c$ and then sample an answer $a_{new}$ by $\text{Bernoulli}(P_{ij})$ for the task $(o_i,o_j)$.

\subsubsection{{\bf Estimating the influence of worker reliability}}% to $a_{new}$}} 

In this section, we discuss how worker reliability influences the predicting process of answer $a_{new}$. As mentioned in Sec.~\ref{sec:predictanswer}, the posterior distribution $P_{ij}$ can be estimated based on the collected answers $A^c$. The estimation framework is built on our underlying assumption that {\em every worker is reliable.} 

We relax this assumption and attempt to add the effect of workers' reliability (i.e., the probability of answering correctly) based on their provided answers in the past. We first define that $rel$ is the average accuracy of answers in $A^c$. Assume that the new task $t_{new}$ is assigned to an unknown coming worker; the probability of answer $a_{new}$ should be revised as follows: 
\begin{equation}\label{eq:reliability}
P'_{ij} = P_{ij} \times rel + (1-P_{ij}) \times (1-rel).
\end{equation}
We use the average reliability of workers already answering $A^c$ to estimate the reliability of coming workers. Note that we can also use the quality of a worker which is recorded in the platforms and calculated by platforms based on workers' answer history in other projects. But we cannot ask platforms the exact quality of a worker when he/she answers our tasks. Instead, we set a lower bound of the quality to filter bad workers and use this lower bound as $rel$ for coming workers.

\subsubsection{{\bf Generating answers in the next batch $A_{i+1}$}}

So far, we have discussed how to predict the next task answer $a_{new}$ based on the collected answers $A^c$ and worker reliability. To predict the answers $A_{i+1}$ obtained in the next batch, we apply an iterative process that generates answers one after another. Algorithm~\ref{alg:predict_remain_ans} shows the pseudo code of the iterative process. We first estimate $P_{ij}$ in line 2-6. Then we utilize the assignment module $\mathcal{T}$ to get the importance of tasks. We select the first $n_\text{batch}$ important tasks, predict the answers respectively and add into $A_{i+1}$ in line 7-12. 

We can also predict a ``complete'' answer set $A$ (obtained when we use up the budget $B$). Based on Algorithm~\ref{alg:predict_remain_ans}, we predict $A_{i+1}$ based on $A^c= A_1\cup ... \cup A_i$. Similarly, $A_{i+2}$ is predicted based on $A_{1}\cup...\cup A_{i+1}$, $A_{i+3}$ is predicted based on $A_{1}\cup...\cup A_{i+2}$ and so on. Finally, we can predict $A = A^c \cup A_{i+1} \cup A_{i+2} \cup A_{i+3}... $ until the size of $A$ is equal to the given budget $B$.

%Given the number of answers $r$ we want to predict, we predict a new answer using previous method, obtain the next answer set $\mathcal{R}$ ($|\mathcal{R}|=r$) and a future result $\sigma_\text{future} = \mathbb{A}(\mathcal{L}_c+\mathcal{R})$. If $r=B-|\mathcal{L}_c|$, it means that we want to know the final result when we used up the budget $B$. If $r$ is large, we need several batches to collect them. When we sample the next answer, we need to add the batch of answers we already predict into the current answer set.

\begin{algorithm}[bt] % \small
 \caption{\emph{Predicting the Next Answer Set}}  \label{alg:predict_remain_ans}
 \KwIn  {Current answer set $A^c$, Inference module $\mathcal{I}$, Task assignment module $\mathcal{T}$, the number of tasks in a batch $n_\text{batch}$}
 \KwOut {Next answer set $A_{i+1}$ }   
 
    {Initialize $A_{i+1} = \varnothing$}\\ 
    
       \vspace{.5em}
       {// \emph{Step 1: Build the matrix $M$ and $P$ }}\\
       {Built matrix $M$ based on the answer in $A^c$}

       \For{all possible $(i,j)$}{
          { Estimate $P_{ij} = \frac{M_{ij}+1}{M_{ij}+M_{ji}+2}$ by Eq.~\ref{eq:pij}}  \\
          { Calculate $P'_{ij}$ by $P_{ij}$ and worker reliability Eq.~\ref{eq:reliability}}
       }

       \vspace{.5em}
       {// \emph{Step 2: Getting the next $n_\text{batch}$ important tasks from $\mathcal{T}$}}\\
       {$T = \mathcal{T}(A^c)$}  \\

    \For { each $t_{new}$ in $T$ } {
       \vspace{.5em}
       {// \emph{Step 3: Predict the answer of $t_{new} =(o_i, o_j)$}}\\
       {Sample $a \sim \text{Bernoulli}(P'_{ij})$ by Eq.~\ref{eq:anew}}\\
       {$A_{i+1}=A_{i+1}\cup \{a\}$}

   }
   
   \Return {$A_{i+1}$;}\\ 

\end{algorithm}

\subsection{Calculating Deviation} \label{sec:calculate_deviation}

In the last section, we showed how to predict a ``complete'' answer set $A$. In this section, we discuss how to judge whether the current point satisfies the definition of the optimal stopping point. %derive the distance from the interim ranked list $\sigma^c = \mathcal{I}(A^c)$ to the final ranking $\sigma = \mathcal{I}(A)$. 

%In section \ref{sec:deviation_definition}, we define the expected distance between the current result and possible final results. Then we analyze the number of required samples to compute the expected distance in section \ref{sec:number_of_samples}. Finally, we describe the total process in section \ref{sec:total_process}.

\subsubsection{{\bf Expected distance between rankings}} \label{sec:deviation_definition}

Given a deterministic answer set $A$, the inference module $\mathcal{I}$ can be used to compute the interim ranking $\sigma_i = \mathcal{I}(A_1\cup...\cup A_i)$ and the distance $\mathbb{D}(\sigma_i,\sigma_j)$ between two rankings (cf. Eq.~\ref{eq:ranklist} and \ref{eq:topklist}). 
%Given the answer sets $A^c$ and $A$, the inference module $\mathcal{I}$ can be used to compute the corresponding ranked lists $\sigma_c$ and $\sigma$ and their distance $\mathbb{D}(\sigma,\sigma_c)$ (cf. Eq.~\ref{eq:ranklist} and \ref{eq:topklist}). 
However, the probabilistic process may create many possible worlds, i.e., many possible answer sets $\mathbb{A}=\{A^1, A^2 ...\}$. If we know the occurrence probability of each possible world $Pr(A')$ where $A' \in \mathbb{A}$, the expected distance between the $i$th and $j$th batches can be defined as 
\begin{equation}
\mathrm{E}[\mathbb{D}_{ij}] = \sum_{A' \in \mathbb{A}} Pr(A') \times \mathbb{D}(\mathcal{I}(A'_1\cup...\cup A'_i),\mathcal{I}(A'_1\cup...\cup A'_j))
\end{equation}

However, it is difficult to calculate the occurrence probability because it is impossible to conduct a brute-force search for all possible worlds. To tackle this problem, we apply the Monte Carlo method, that allows an estimation of the sampling distribution of almost any statistic using random sampling method. The Monte Carlo method helps to generate a list of possible worlds, i.e., ``complete'' answer sets $\{A^1, A^2, ..., A^s,...| s \in [1,n_\text{sample}]\}$. Given a pair $(i,j)$, we are able to compute a list of pairs of rankings $(\sigma_i^s$, $\sigma_j^s)$ and the corresponding distances $\mathbb{D}_{ij}^s$. By the law of large numbers, the expected distance $\mathrm{E}[\mathbb{D}_{ij}]$ can be approximated by taking the sample mean 
\begin{equation} \label{eq:sample_mean}
\overline{\mathbb{D}}_{ij} = \frac{1}{n_\text{sample}}\sum_{s=1}^{n_\text{sample}} \mathbb{D}(\mathcal{I}(A_1^s\cup...\cup A_i^s),\mathcal{I}(A_1^s\cup...\cup A_j^s)).
\end{equation}
If $p_\text{current}$ is the earliest point satisfying $\forall p_i,p_j \in [p_\text{current}, p_\text{final}]$, $\overline{\mathbb{D}}_{ij} \leq \theta$, $p_\text{current}$ is the stopping point decided by our ES module.

\subsubsection{{\bf The number of required samples}} \label{sec:number_of_samples}

In the Monte Carlo method, it is important to decide the number of required samples such that the quality is secured. Following common practice, we use Hoeffding's inequality\cite{hoeffding} to decide it. 

\stitle{Hoeffding's Inequality} Let $X_1,...,X_n$ be independent random variables bounded by the interval $[0,1]: 0 \leq X_i \leq 1$. Define the mean of these variables as $\overline{X} = \frac{1}{n}(X_1+...+X_n)$. Then
\begin{equation}
Pr(\mathrm{E}[\overline{X}] -\overline{ X }\geq t) \leq e^{-2nt^2}
\end{equation}
where $t \geq 0$. 

We regard a possible world answer set $A^s$ as a sample. The distance $\mathbb{D}_{ij}^s$ can be regarded as an independent random variable given $(i,j)$. Based on Hoeffding's Inequality, we have $Pr( \mathrm{E}[\mathbb{D}_{ij}] -\overline{\mathbb{D}}_{ij} \geq t) \leq e^{-2nt^2}$. 
This inequality could be transformed into a confidence interval of $\mathrm{E}[\mathbb{D}_{ij}]$:
\begin{equation}
Pr( \mathrm{E}[\mathbb{D}_{ij}] \le \overline{\mathbb{D}}_{ij} + t) > 1-e^{-2nt^2},
\end{equation}
where $\overline{\mathbb{D}}_{ij}$ is computed using Eq.~\ref{eq:sample_mean}. We require at least $\frac{\text{ln}(1/\alpha)}{2t^2}$ samples to acquire $(1-\alpha)$-confidence interval for $\mathrm{E}[\mathbb{D}_{ij}] \le \overline{\mathbb{D}}_{ij} + t$. 

Given the targeted accuracy tolerance $\theta$, if we find that $\overline{\mathbb{D}}_{ij} \le \theta - t$, we can also derive
$Pr( \mathrm{E}[\mathbb{D}_{ij}] \le \overline{\mathbb{D}}_{ij} + t \le \theta ) > 1- e^{-2nt^2}.$
We summarize it as the following theorem.
\begin{theorem}
Given two points $p_i$ and $p_j$, we secure that $\mathrm{E}[\mathbb{D}_{ij}] \le \theta$ with confidence $(1-\alpha)$ after we random sample $\frac{\text{ln}(1/\alpha)}{2t^2}$ ``complete'' answer sets and find $\overline{\mathbb{D}}_{ij} \le \theta - t$, for some $0 < t < \theta$.
\end{theorem}

Here we set the confidence level $\alpha = 5\%$ and the estimation error $t$ as an order of magnitude smaller than $\theta$ which secures enough samples to give a good estimation.
We need to sample $n_\text{sample} \approx 10^{4}$ for $\theta = 0.1$, and $n_\text{sample} \approx 10^{6}$ when we set $\theta = 0.01$. %Multi-thread or distributed computation can be used to accelerate the process, as it is natural for parallel processing. 
The workload of sampling can be accelerated by multithreading or distributed computation.

We then analyze the number of samples to secure all $\mathrm{E}[\mathbb{D}_{ij}] \le \theta$ with high probability from the current to the final state, i.e., judge whether the following formula holds:
$\forall p_i,p_j \in [p_\text{current}, p_\text{final}]$, $~\mathrm{E}[\mathbb{D}_{ij}] \leq \theta.$

Assume that number of batches for remaining budget is $m = \frac{B-|A^c|}{n_\text{batch}}$, there are $(m+1)m/2$ different expected distances needed to compute and check. 
If we acquire confidence $(1-\alpha')$ for all the expected distances, the confidence $(1-\alpha)$ and the number of samples for each expected distance can be set as:
% \begin{equation} \label{eq:number_of_samples}
% \alpha = \frac{\alpha'}{(m+1)m/2}~~ \text{and} ~~n_\text{sample}= \frac{\text{ln}((m+1)m/2)+\text{ln}(1/\alpha')}{2t^2}.
% \end{equation}

%\begin{equation} 
\begin{gather}\label{eq:number_of_samples}
\alpha  = \frac{\alpha'}{(m+1)m/2}~~ \text{and}  \\
n_\text{sample} = \frac{\text{ln}((m+1)m/2)+\text{ln}(1/\alpha')}{2t^2} .
\end{gather}
%\end{equation}

We utilize the union bound to prove Eq.~\ref{eq:number_of_samples}. 
%The confidence $(1-\alpha')$ means that $\exists p_i,p_j, \mathrm{E}[\mathbb{D}_{ij}] > \overline{\mathbb{D}}_{ij} + t$ with $\alpha'$ probability. 
Let $\mathrm{E}[\mathbb{D}_{ij}] \le \overline{\mathbb{D}}_{ij} + t$ for a pair $(p_i,p_j)$ be an event.
The confidence $(1-\alpha)$ means the probability that one event fails is $\alpha$. %while the confidence $(1-\alpha')$ means the probability that at least one of the events fails is $\alpha'$. 
Then based on the union bound, we derive that the probability that at least one of the events fails is no greater than the sum of the probabilities of the individual events, which is $\sum_{i=1}^{(m+1)m/2} \alpha = \alpha'$ using Eq.~\ref{eq:number_of_samples}. In other words, the probability that no event fails is at least $\alpha'$, which satisfies our requirement.

% one simulated collection of remaining answer set $\mathcal{R}$ and obtaining a possible final rank $\sigma_\text{final}$ as a sample. The distances $\mathbb{D}(\sigma_\text{curr},\sigma_\text{final})$ could be regarded as independent random variables. Based on Hoeffding's Inequality, we have $Pr( \mathrm{E}[\mathbb{D}] -\overline{ \mathbb{D} }\geq t) \leq e^{-2nt^2}$.
% This inequality could be transformed into a confidence interval of $\mathrm{E}[\mathbb{D}]$:
% $$
% Pr( \mathrm{E}[\mathbb{D}] \le \overline{\mathbb{D}} + t) > e^{-2nt^2}
% $$
% We require at least $\frac{\text{ln}(1/\alpha)}{2t^2}$ samples to acquire $(1-\alpha)$-confidence interval for $\mathrm{E}[\mathbb{D}] \le \overline{\mathbb{D}} + t$.
% If we ensure $\overline{\mathbb{D}} \le t_\text{acc} - t$, we have 
% $$Pr( \mathrm{E}[\mathbb{D}] \le \overline{\mathbb{D}} + t \le t_\text{acc}) > e^{-2nt^2}.$$

\subsection{{\bf Putting it all together}} \label{sec:total_process}
%In Sec.~\ref{sec:predictanswer} and Sec.~\ref{sec:number_of_samples}, we showed how to predict the next answer $A_{i+1}$ and how estimate the expected distance from the current answer set $A^c$ to the final state $\mathbb{A}$. 
In this section, we put all of these techniques together to finalize the ES module, as shown in Algorithm~\ref{alg:judge_point}.
We first calculate how many batches we need to predict the remaining budget (in line~\ref{alg:line:batch}) and estimate the number of needed samples by Hoeffding's Inequality (in line~\ref{alg:line:hoeffding}).
During each sample, we use our probabilistic model in Sec.~\ref{sec:predictanswer} to predict a new batch of answers $A_j$ and repeat $m$ times to obtain a complete answer set (in lines~\ref{alg:line:sample:begin}-\ref{alg:line:sample:end}). The temporary answer set $A$ is current answers we ``collect'' including actual answers $A^c$ and predicted answers. 
We record the inferred ranked list $\sigma[j]$ after predicting the $j$th batch of answers. At the end of $s$th sample, we know all the ranked lists and compute the distance $d[s][i][j]$ between the $i$th and $j$th batches. 

We then calculate $\overline{\mathbb{D}}[i][j]$ as the expected distance $E[\mathbb{D}_{ij}]$ for each remaining batch (in lines~\ref{alg:line:avgdist}-\ref{alg:line:avgdist:end}), which is the mean of sampled distances between the interim ranked lists $\sigma[i]$ and $\sigma[j]$ in the stopping points $p_i$ and $p_j$.
We invoke a programming call to terminate the rank process $\mathcal{R}$ when the expected distances fulfill the accuracy tolerance $\theta$ (in lines \ref{alg:line:tolerance}-\ref{alg:line:tolerance:end}).

\begin{algorithm}[t!] %\small
    \caption{\emph{Early-Stopping module}}  \label{alg:judge_point}
 \KwIn  {Current answer set $A^c$, Inference module $\mathcal{I}$, Distance function $\mathbb{D}$, Budget $B$, accuracy tolerance $\theta$, confidence interval $\alpha'$}
 %, distance array $d[][][]$ and $\overline{\mathbb{D}}[][]$}
  
 %{Initialize $\sigma^c = \mathcal{I}(A^c)$} by the inference module\\
 {Calculate number of batches for remaining budget $m=\frac{B-|A^c|}{n_\text{batch}} $}\label{alg:line:batch}\\
 {Estimate the number of samples $n_\text{sample}$ by Eq.~\ref{eq:number_of_samples}}\label{alg:line:hoeffding}\\
 {Initialize distance array $d$ and $\overline{\mathbb{D}}$}

    \For{$1\le s \le n_\text{sample}$}{ \label{alg:line:sample:begin}
       {Set a temporary answer set $A = A^c$} \\
       {Create a temporary array of ranked lists $\sigma$ and set $\sigma[0] =  \mathcal{I}(A^c)$} \\
       \For{$1\le j \le m $}
       { 
          {Predict a batch of answers $A_j$ based on $A$ by Alg.~\ref{alg:predict_remain_ans}} \\ %\label{alg:line:dist:start}}\\
          {$A = A \cup A_j $} \\
          {$\sigma[j] = \mathcal{I}(A)$} \\
        }
        \For{$0\le i \le m-1 $}{
          \For {$ i+1 \le j \le m $}{
            {$d[s][i][j] = \mathbb{D}(\sigma[i],\sigma[j])$} %\label{alg:line:dist:end}} \\
            } 
        }
    }\label{alg:line:sample:end}
    \For{$0\le i \le m-1 $\label{alg:line:avgdist}}{
       \For{$i+1\le j\le m $ } {
            {$\overline{\mathbb{D}}[i][j] = \frac{1}{n_\text{sample}} \sum_{1 \le s \le n_\text{sample} } d[s][i][j] $} \\
       }   
    }\label{alg:line:avgdist:end}
 
    \If{ $\overline{\mathbb{D}}[i][j] \leq \theta - t, \forall i,j $ \label{alg:line:tolerance}} 
    {Invoke a programming call to terminate the rank process $\mathcal{R}$} \label{alg:line:tolerance:end}
    \Else {Continue collecting the next batch of answers}
 \end{algorithm}

%!TEX root = main.tex

\section{Experimental Evaluation}\label{sec:experiments}
\setlist{leftmargin=3mm}

In this section, we thoroughly evaluate our ES module on two real public datasets, which are already collected by others in AMT. Based on the different inference algorithms and task assignment approaches, we compare our ES module with some standard quality estimation methods. 
%We evaluate the effectiveness of our stopping criterion using two real datasets, by comparing it to alternative stopping methods. First, we present the experimental settings (Section \ref{sec:experimental_setting}); then, we assess the performance of different stopping criteria for ranking and top-$k$ list computation by varying the acceptable error and the combinations of aggregation and assignment methods (Section \ref{sec:experimental_results}). We also vary different parameters, such as $k$, $B$, and $n_\text{batch}$ to test the robustness of our method in section \ref{sec:experimental_analysis}. %Finally we show the effectiveness of our early-stopping module in practice in section \ref{sec:case_study}.
The source code and datasets used can be found in \url{https://www.dropbox.com/sh/lp0vu57h13oir33/AACSCYZPOcV33bJSFW6bx85ca?dl=0}.

\subsection{Experimental Settings} \label{sec:experimental_setting}

\noindent {\bf Datasets.} We use two real public datasets collected in AMT.
\begin{itemize}
  \item {\it PeopleNum}~\cite{datasetPeopleNum} concerns 39 images taken in a mall, each of which includes multiple persons. The goal is to find the images with the most people in them. 6066 answers were collected from 197 workers. Each pair of images is answered by at least 5 workers. 
  \item {\it PeopleAge}~\cite{exptopk}  has 50 human photos with ages from 50 to 100. The goal is to find the photos that include the youngest person. There are 4930 answers from 150 workers. Each pair of photos is answered 3 times at least. 
%  \item {\bf ImageClarity}\cite{exptopk}. It contains 100 images with different clarities. The goal is to judge clarity of images. It has 25955 answers and no worker information. Most of pairs are answered by 5 times and 6\% pairs are never being answered. 
\end{itemize}
%These datasets have different scales and levels of difficulties.
%\nikos{what is the difference between scale and level? Can you simply put level of difficulty here? I think the scale is similar in both datasets, if scale is the size}
%PeopleAge is a more subjective dataset as it is about to decide the people age in photos. The variance of answers is higher than that of ImageClarity. Another reason is that the workers in ImageClarity are motivated by clarity. 
{\it PeopleAge} is hard because it is relatively subjective and different workers may have different opinions on age. The difficulty of {\it PeopleNum} is medium because it costs some time to count the persons. 

%{\bf ImageClarity} is more objective and the overall answer quality is good since the workers are motivated by clarity.  

%because it is relatively subjective and different workers might have different opinions on age. The scale of PeopleAge is medium. 
%{\bf ImageClarity} is the largest with 100 items and the easiest since clarity is objective and quick to judge. 
%{\bf PeopleNum} has a small size and a medium level of difficulty because it costs some time to count the persons in the photos. 
% \caihua{
%   {\bf PeopleAge} is hard because it is relatively subjective and different workers might have different opinions on age.
%   The difficulty of {\bf PeopleNum} is medium because it costs some time to count the persons in the photos. 
% }

\stitle{Inference Modules $\mathcal{I}$} According to~\cite{exptopk}, we select some recommended inference algorithms and task assignment strategies to work with our ES module. For rank inference algorithms, we choose 4 methods: %{\it Copeland} \cite{Copeland}, {\it  Iterative} \cite{local}, {\it Local} \cite{local} and {\it CrowdBT} \cite{crowdbt}. 
\begin{itemize}
  \item{\it Copeland} is a basic election approach where the objects are sorted by the times they win/lose in the comparisons. 
  \item{\it Local} is a heuristic-based method based on a comparison graph, where nodes are objects and edges are built based on the pairwise comparisons. The score of an object is defined by the number of winning objects minus the number of losing objects in its 1-hop and 2-hop neighborhood.  
  \item{\it Iterative} is an extended version of {\it local} supporting top-$k$ queries. It keeps discarding the bottom half of the objects in the inference process and then re-computes the scores of the surviving objects. It repeats these two processes until $k$ objects are left. 
  \item{\it CrowdBT} is a representative method that uses the Bradley-Terry (BT) model to estimate the latent score $s_i$ of the object $o_i$. It models the probability $o_i \prec o_j$ as $\frac{e^{s_i}}{e^{s_i}+e^{s_j}}$. Based on the crowdsourced comparisons $A$, it computes scores for the objects by maximizing $\sum_{o_i \prec o_j \in A} \text{log}(\frac{e^{s_i}}{e^{s_i}+e^{s_j}})$.
\end{itemize}

\stitle{Task Assignment Modules $\mathcal{T}$} We implemented 4 task assignment strategies based on commercial systems and existing work. 

\begin{itemize}
  \item{\it Random} is the strategy used by Amazon Mturk; tasks are assigned to coming workers at random and all tasks are answered the same number of times.  
  \item{\it Greedy} chooses the pair of objects with the highest product of scores as the next task. 
  \item{\it Complete} finds the top-$x$ objects with the highest scores, where $x$ is the largest integer satisfying $\frac{x(x-1)}{2} \leq n_\text{batch}$, and sets their pairwise comparisons as the next tasks.
  \item{\it CrowdBT} is an active learning method which selects the pair of objects which maximizes the information gain based on the estimated scores. 
\end{itemize}

%   We regard the strategy used by Amazon Mturk as {\it Random}; the task for the coming worker is assigned at random and there is a guarantee that all tasks are answered the same number of times. 
% {\it Greedy} and {\it Complete} are heuristic-based methods that select the objects to assign which maximize the probability of obtaining the top-$k$ results early. 
%{\it Greedy} chooses the pair of objects with the highest product of scores. 
% {\it Complete}, finds the maximum number of objects depending on $n_\text{batch}$ and makes a pairwise comparison on them. \nikos{unclear which objects are selected at each batch} 
% \caihua{{\it Complete} finds the top-$x$ objects with highest scores, where $x$ is the largest integer satisfying $\frac{x(x-1)}{2} \leq n_\text{batch}$, and makes a pairwise comparison on them. }
% {\it CrowdBT} is an active learning method which selects the objects for the next comparison that maximize the information gain based on the estimated scores. 
% Active learning methods achieve high quality but have low efficiency. Different assignment strategies should be chosen in different situations. For example, {\it Random} is acceptable if we have a large budget which allows each pair of objects to be compared by multiple workers. {\it CrowdBT} is a good choice if we emphasize on the quality and do not bother about the required time.
%have an online requirement.
  
Based on the characteristics of the inference algorithms and the task assignment strategies, we form and test 7 rank processes $\mathcal{R}$:
%inference algorithm and task assignment strategy: 
{\it Copeland-Random}, {\it Iterative-Random}, {\it Local-Random}, {\it CrowdBT-Random}, {\it Local-Greedy}, {\it Local-Complete}, and {\it CrowdBT-CrowdBT}. 

\noindent {\bf Competitors.} In order to evaluate our ES module, we also investigate two alternative stopping criteria based on statistical analysis.
% implement our proposal {\it Early-Stopping (ES) module} (cf. Sec.~\ref{sec:approach}) and three alternative early stopping criteria.
%three early competitors to our approach. 
\begin{itemize}
%  \item Fix Point: Stop when we consumes certain percentage of budget. We try all the possible percentages and use the best one. It is an oracle method and we use it as a reference.
%  \item {\bf Tournament}. Stop when percentage of pairs of items has been asked a certain time at the first time. The number of asked times could be 1,2,3 and percentage of pairs is 50\% to 100\% at interval of 5\%. For instance, (2,80\%) is one parameter which finds the first point satisfied 80\% pairs of items has been asked twice.
  \item {\it Moving Average (MA)} stops when the following equation is smaller than the threshold $\theta$ at the first time. We calculate the distances between all pairs of consecutive rankings or top-$k$ lists, generated at the last $w$ points before the current stage and average them. Suppose we already collected $i$ batches of answers: 
  \begin{equation} \small
  \text{MA}(i,w) = \frac{ \sum_{j=1}^{w} \mathbb{D}(\mathcal{I}(A_1 \cup ... \cup A_{i-j}), \mathcal{I}(A_1 \cup ... \cup A_{i-j+1})) }{ w}
  \end{equation}

  \item {\it Weighted Moving Average (WMA)} is similar to MA, %we calculate the distance between two consecutive rankings or top-$k$ lists generated by previous $w$ points from current point $p$.
  except that we assign different weights to the distances based on how far away they are from the current stage.
  % for the distance.
  The distance between the latest two rankings has the largest weight $w$, the second latest $w-1$, etc, and so on. 
  \begin{equation} \scriptsize
  \text{WMA}(i,w) = \frac{ \sum_{j=1}^{w} (w-j+1) \mathbb{D}(\mathcal{I}(A_1 ... A_{i-j}), \mathcal{I}(A_1 ... A_{i-j+1})) }{ w(w+1)/2}
  \end{equation}

\end{itemize}

\stitle{Evaluation Metrics} 
We define the optimal stopping point $p_\text{optimal}$ in the Sec.~\ref{sec:stopping_point}.
%We simulate the whole collection process until the budget is exhausted so that we know the best stopping point $p_{\text{best}}$ defined in \ref{def:best_stopping_point} 
% $$
% \forall i \in [p_{\text{best}}, B/n_\text{batch}], \quad \mathbb{D}(\sigma_i,\sigma_\text{final}) \leq \theta
% $$
% where $\sigma_{i} = \mathcal{I}(A_1 \cup ... \cup A_i)$ is the ranked list after we have collected the $i$-th batch of answers and $\sigma_\text{final} = \mathcal{I}(A)$ is the ranking computed from the complete answer set $A=(A_1 \cup ... \cup A_{B/{n_\text{batch}}})$. Without loss of generality, we assume $B$ is divided evenly by $n_\text{batch}$ and $B/{n_\text{batch}}$ is the number of batches we need to collect the total answers. A point $p$ can be viewed as the $p$-th batch, which is also regarded as the number of answers collected already. For any point between $p_\text{best}$ and the final state $B/n_\text{batch}$, the distance should not exceed the accuracy tolerance $\theta$. 
To evaluate the effectiveness of different stopping criteria, we analyze the difference between $p_\text{optimal}$ and the stopping point $p_\text{sc}$ predicted by a stopping criterion. Mathematically:  
\begin{equation} \label{eq:delta_sc}
\Delta_\text{sc} = \frac{|p_{\text{optimal}}-p_{\text{sc}}|}{B/n_\text{batch}}
\end{equation}

However, $\Delta_\text{sc}$ cannot reveal all performance factors of different early-stopping strategies. The stopping point is actually a trade-off between the budget used and the accuracy. If $p_\text{sc}$ stops ahead of $p_\text{optimal}$, then this strategy saves more budget but loses accuracy (as the ranked list will be inferred based on fewer answers). To better reveal the real performance of different strategies, we also calculate the percentage of the used budget and the actual error, where 
\begin{equation}
\begin{aligned}
&\text{Used Budget} = \frac{p_\text{sc}}{B/n_\text{batch}} \text{~,~} \\
&\text{Actual Error} = \mathbb{D}(\sigma_{p_\text{sc}},\sigma_\text{final})
\end{aligned}
\end{equation}

% cannot indicate whether $p_{\text{SC}}$ is smaller or larger than $p_{\text{BSP}}$. If $p_{\text{SC}}$ is larger than $p_{\text{BST}}$, SC wastes some budget to collect more answers while the expected distance must be smaller than $\theta$ when SC stops. If $p_{\text{SC}}$ is smaller than $p_{\text{BSP}}$, SC saves some budget but the distance between the result at $p_{\text{SC}}$ and the final result may be larger than $\theta$, which violates the requester's requirements. Thus we also calculate the percentage of used budget and actual distance for a stopping criterion SC, where 

% \begin{figure*}[ht]
%  \centering
%    \includegraphics[width=1.0\textwidth]{figs/noselect_observation.eps}
%   \caption{Observation: Quality vs. Budget (without Assignment Strategies)}
%   \label{fig:noselect_observation} 
% \end{figure*}

% \begin{figure*}[ht]
%  \centering
%    \includegraphics[width=1.0\textwidth]{figs/addselect_observation.eps}
%   \caption{Observation: Quality vs. Budget (with Assignment Strategies)}
%   \label{fig:addselect_observation} 
% \end{figure*}

\subsection{Experimental Results} \label{sec:experimental_results}

\begin{figure*}[ht]
 \centering
   \includegraphics[width=1\textwidth]{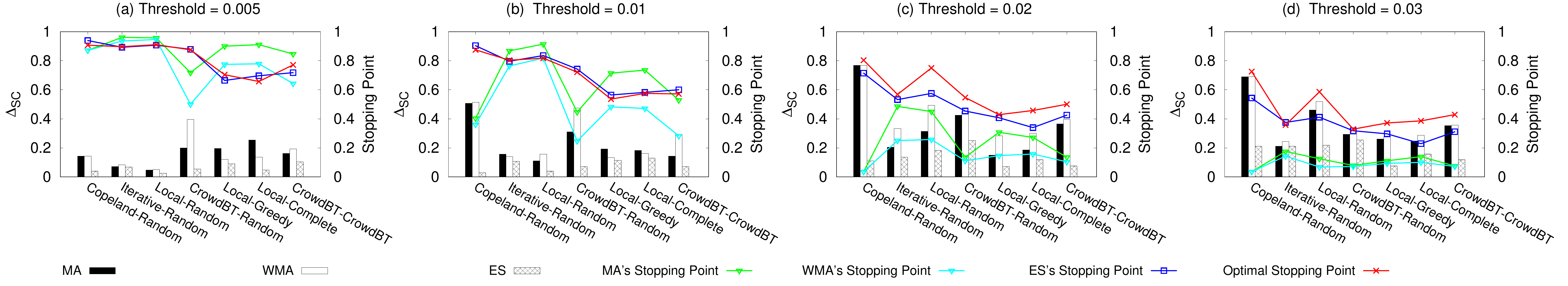}
  \caption{$\Delta_{\text{sc}}$ \& Stopping Points in PeopleNum Dataset for Top-$10$ Lists}
  \label{fig:topk_peopleNum_result} 
\end{figure*}

\begin{figure*}[ht]
 \centering
   \includegraphics[width=1\textwidth]{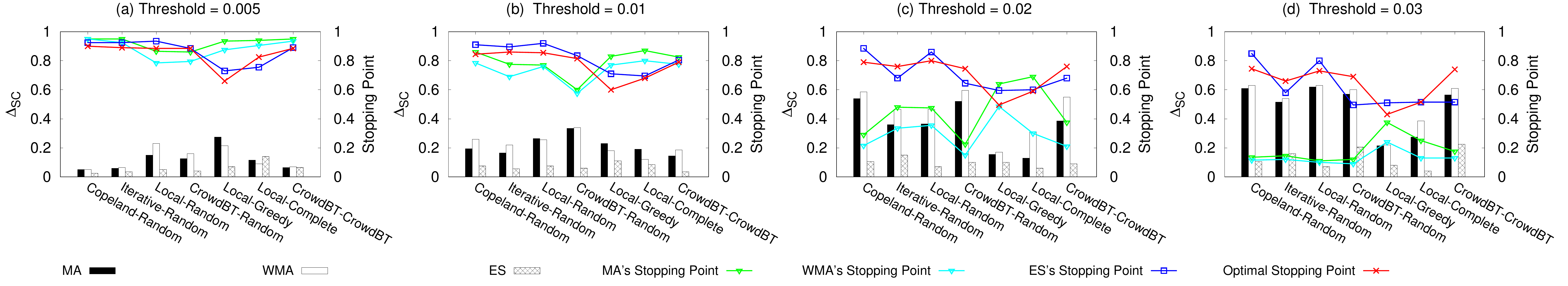}
  \caption{$\Delta_{\text{sc}}$ \& Stopping Points in PeopleAge Dataset for Top-$10$ Lists}
  \label{fig:topk_peopleAge_result} 
\end{figure*}

\begin{figure*}[ht]
 \centering
   \includegraphics[width=1\textwidth]{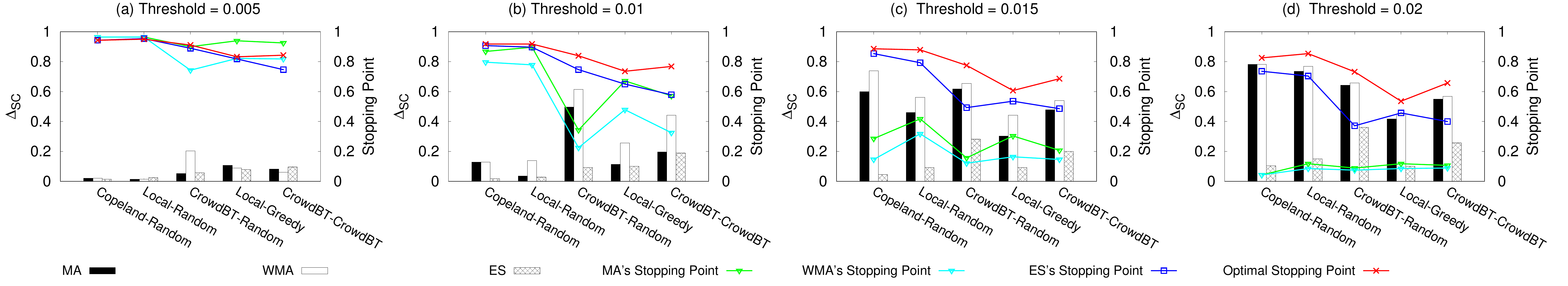}
  \caption{$\Delta_{\text{sc}}$ \& Stopping Points in PeopleNum Dataset for Rankings}
  \label{fig:rank_peopleNum_result} 
\end{figure*}

\begin{figure*}[ht]
 \centering
   \includegraphics[width=1\textwidth]{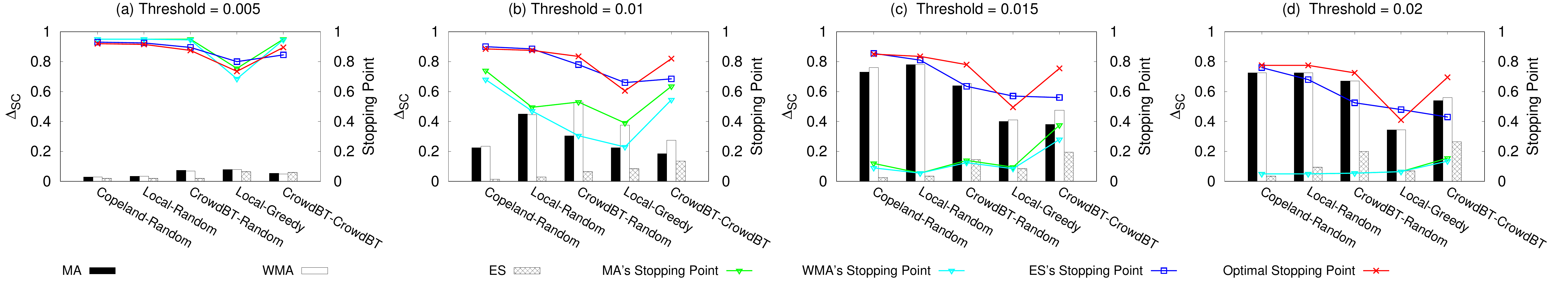}
  \caption{$\Delta_{\text{sc}}$ \& Stopping Points in PeopleAge Dataset for Rankings}
  \label{fig:rank_peopleAge_result} 
\end{figure*}

\subsubsection{\bf Implementation details}

We compare our ES module with two competitors, MA and Weighted MA, on two datasets for ranking or top-$k$ queries. The objective is to show the superiority and robustness of ES on top of different rank processes $\mathcal{R}$.%inference algorithms and task assignment strategies. 

% The aim is to demonstrate that the performance of AnsBoost is consistently better than others in the different situations, no matter what kind of aggregation algorithms, assignment strategies and requesters' requirements (pre-set budget and acceptable error $t_{\text{acc}}$) we use. We simulate the whole collection process based on the answers in the real datasets. To solve the cold-start problem and make each assignment method workable, we allow each micro-task to have one answer at first and then collect answers for micro-tasks in a batch decided by the assignment method.

The total budget is set to the number of answers in each original dataset. The number of microtasks in a batch is set to 200. To get an answer of a microtask $(o_i,o_j)$, we randomly sample an answer from the answer set of $(o_i,o_j)$ without replacement. If some pairs are running out of answers, we will simulate the next answer by a worker that has average reliability. To solve the cold-start problem of some task assignment strategies, we pre-generate an answer to every pairwise comparison (microtask). 

We also choose the best window size for MA and Weighted MA, which is 20 for PeopleNum dataset and 10 for PeopleAge dataset, respectively. Besides, a little change of initial answers for the cold-start problem will change the next sequence of microtasks. Thus, we run the collection process 10 times, and report the average performance.

% Between each batch, we apply for different stopping criteria to decide whether to stop. We use the original number of answers in PeopleNum and PeopleAge datasets as the pre-set budget. In spite of that, there may not be enough answers for a micro-task in original datasets because an assignment method asks it several times. We use an extra worker with average reliability to complement missing answers. We choose the best window size of MA and Weighted MA, which is 20 for PeopleNum dataset and 10 for PeopleAge dataset. Besides, a little change of initial answers for the cold-start problem will change the next sequence of micro-tasks. Thus, we run the collection process 20 times, and the performance is the average of them. 

We use two y-axes in Fig.~\ref{fig:topk_peopleNum_result} - \ref{fig:rank_peopleAge_result}. The left y-axis is $\Delta_{\text{sc}}$, which is defined in Eq.~\ref{eq:delta_sc}. The right y-axis is the relative stopping point of MA, Weighted MA, ES and the optimal stopping point, which divided by the maximum possible stopping point, $B/n_\text{batch}$. Table~\ref{table:avg_saved_budget_topk_list} and \ref{table:avg_saved_budget_ranking} show the used budget and actual error for the optimal stopping point and the stopping point decided by ES and Weighted MA averaging by the seven rank processes $\mathcal{R}$.

\subsubsection{{\bf Top-$k$ ranking}}

Fig.~\ref{fig:topk_peopleNum_result} and \ref{fig:topk_peopleAge_result} show $\Delta_\text{sc}$ and the stopping point of our ES module and the two alternative stopping criteria for top-$k$ queries. Each stopping criterion is evaluated with seven rank processes (cf. Sec.~\ref{sec:experimental_setting}). We set $k=10$ by default. The accuracy tolerance $\theta$ is set to $\{0.005$, $0.01$, $0.02$, $0.03\}$. For instance, $\theta=0.01$ means the possible number of inverse pairs between the current ranked list and the final state is smaller than $10^2 \times 0.01 = 1$.

ES outperforms the other two competitors for different rank processes and different datasets in all settings except for {\it Local-Complete} when $\theta= 0.005$ on PeopleAge dataset. When we set $\theta$ to a larger value (accepting higher accuracy loss), MA and Weighted MA tend to fall into the false states mentioned in the Sec.~\ref{sec:stopping_point} and stop much earlier than the optimal stopping point, which results in high accuracy loss.

% are for top-$k$ lists. We set k as 10. We try $t_{\text{acc}} = 0.005, 0.01, 0.02$ and $0.03$ for seven different combinations of aggregation and assignment methods. The meaning of thresholds is the number of maximum swaps in the current and the final top-$k$ list. If k is equal to 10, the number of maximum swaps is $t_{\text{acc}} \times k^2$, which is from 0 to 3. 

% With the different thresholds and datasets, $\Delta_{\text{SC}}$ of AnsBoost is always smaller than the value of MA and Weighted MA except when we use {\it Local-Complete} and $t_{\text{acc}}= 0.005$ in PeopleAge Dataset. However, $\Delta_{\text{SC}}$ are both small in that case, which means both methods perform well. 
% MA and Weighted MA perform badly and stop very early when $t_{\text{acc}}=0.02$ or $0.03$. The reason is the distance among different batches is smaller than the threshold at the very beginning which is hard to represent the gap between the current result and the final result. Our method still performs acceptably. 
% In the general situation, AnsBoost has dominated advantages compared with MA and Weighted MA. 

According to the right y-axis, the position of $p_\text{optimal}$ varies from 0.5 to 0.9. The stopping point of our ES module is very close to $p_\text{optimal}$ when compared with the stopping points of MA and Weighted MA. This reveals that ES is effective in finding $p_\text{optimal}$. 
% The curves of AnsBoost and the BSP are almost the same when $t_{\text{acc}} = 0.005$ and $0.01$ but a little bit far away when $t_{\text{acc}}=0.02$ and $0.03$. The reason is small thresholds make we collect more answers and have enough information to predict accurately.

Table~\ref{table:avg_saved_budget_topk_list} shows the average used budget and the actual error of the optimal stopping point, ES and Weighted MA. 
Weighted MA always stops earlier rendering the actual error larger than desired. The maximum actual error of Weighted MA is more than 3 times larger than $\theta$, while ES has an error less than 1 times larger than $\theta$.
% The average actual error of AnsBoost is smaller than $t_{\text{acc}}$ when $t_{\text{acc}} = 0.005$ and $0.01$ and is a little bit larger than $t_{\text{acc}}$ when $t_{\text{acc}} = 0.02$ and $0.03$. The maximum actual error of AnsBoost is $\textbf{0.x}$ greater than $t_\text{acc}$.

\subsubsection{{\bf Complete Ranking}}

Fig.~\ref{fig:rank_peopleNum_result} and \ref{fig:rank_peopleAge_result} show the performance for ranking queries. The accuracy tolerance $\theta$ is set to $\{0.005$, $0.01$, $0.15$, $0.02\}$. Note that we exclude two inference algorithms, {\it Iterative} and {\it Complete}, since they are designed for top-$k$ queries. 

% for the ranking problem. We try $t_{\text{acc}} = 0.005, 0.01, 0.015$ and $0.02$ for aggregation and assignment methods except {\it Iterative-Random} and {\it Local-Complete} because {\it Iterative} and {\it Complete} are designed for top-$k$ list problems and they need $k$ as an input. The meaning of thresholds is the number of maximum swaps in the current and final ranking. However, the denominator of the distance function for ranking is $n(n-1)/2$, so the actual number of swaps is 3.7, 7.4, 11.1 or 14.8 in PeopleNum dataset and 6.1, 12.2, 18.4 and 24.4 in PeopleAge dataset. 

Similar to top-$k$ queries, our ES module is much better than the other two competitions in terms of $\Delta_{sc}$. The curve of ES's stopping point is very close to that of $p_\text{optimal}$ compared with MA and Weighted MA. 
% No matter in which situation, $\Delta_{\text{SC}}$ of AnsBoost is smaller than or similar with $\Delta_{\text{SC}}$ of MA and Weighted MA.
%When $\Delta_{\text{SC}}$ of AnsBoost is larger than the value of MA or Weighted MA, their $\Delta_{\text{SC}}$ are both small and the difference between their $\Delta_{\text{SC}}$ is also tiny, which means their performances are both great in that situation. Similarly, the curve of AnsBoost's stopping point is closer to the BSP compared with MA and Weighted MA. 
In Table~\ref{table:avg_saved_budget_ranking}, the maximum actual error of Weighted MA is more than 4 times larger than $\theta$ while the maximum actual error of ES is less than 2 times larger than $\theta$.

\subsection{Parameter Analysis} \label{sec:experimental_analysis}

In this section, we test the effect of some parameters, including $k$ in top-$k$ queries, the total budget $B$ and the number of microtasks in one batch $n_\text{bach}$. We evaluate these parameters with two rank processes, {\it Local-Random} and {\it Local-Greedy} on PeopleNum dataset.

\begin{figure*}[ht!]
\centering
 \begin{minipage}[t]{0.49\textwidth}
   \includegraphics[width=1\textwidth]{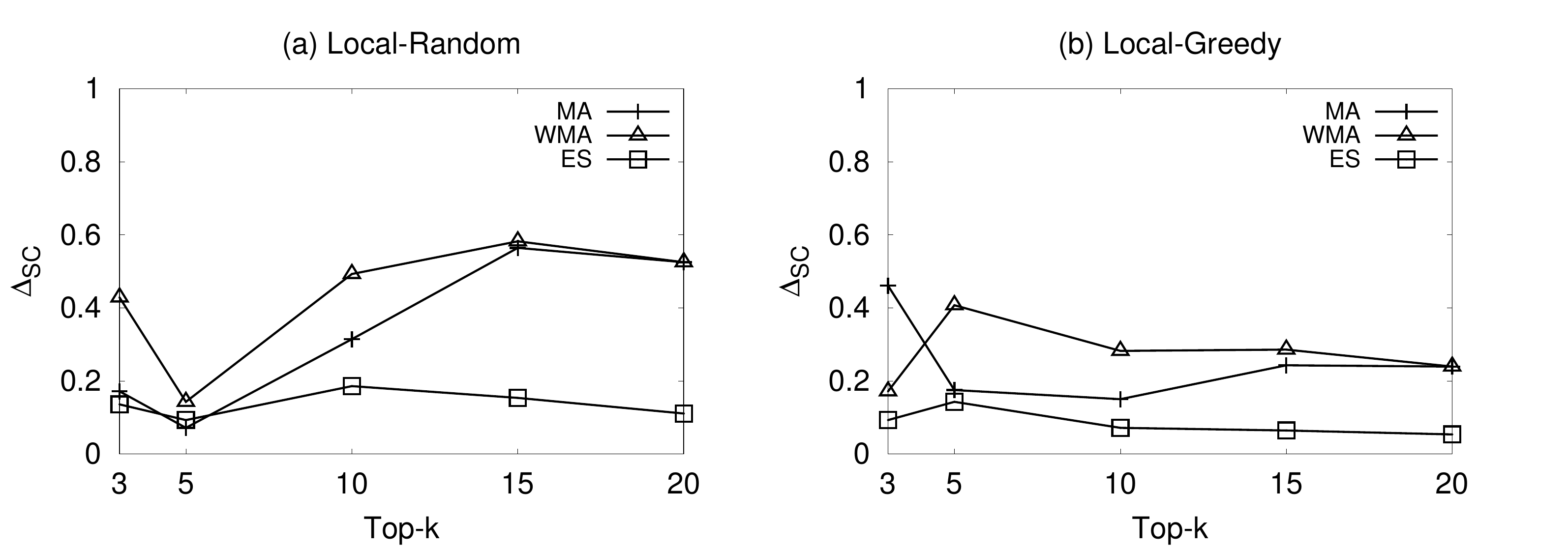}
  \caption{$\Delta_{\text{sc}}$ in varied $k$}
  \label{fig:diff_topk} 
\end{minipage}
\begin{minipage}[t]{0.49\textwidth}
 \centering
   \includegraphics[width=1\textwidth]{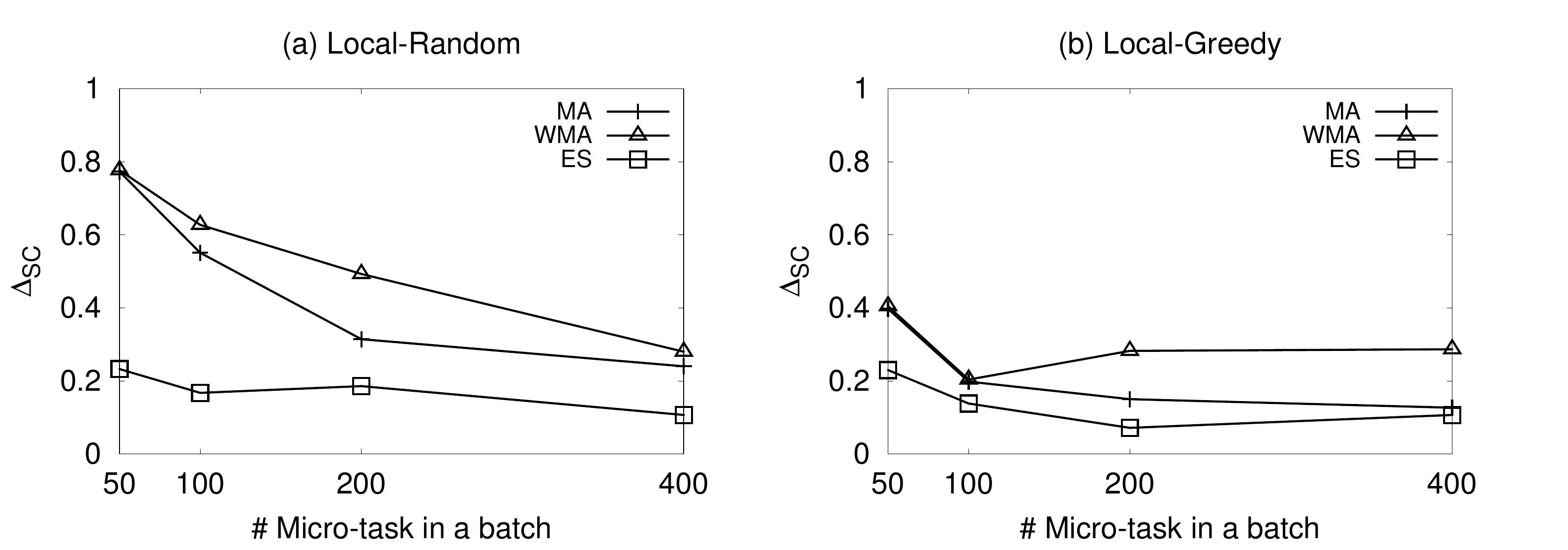}
  \caption{$\Delta_{\text{sc}}$ in varied $n_\text{batch}$}
  \label{fig:diff_n_batch} 
\end{minipage}
\end{figure*}

Fig.~\ref{fig:diff_topk} shows the effect of $k$ in top-$k$ queries and Fig.~\ref{fig:diff_n_batch} shows the effect of $n_\text{batch}$. In these experiments, we set the budget $B$ equal to the total number of answers in the original dataset and set $\theta=0.02$. ES is the clear winner since its $\Delta_\text{sc}$ is always less than or equal to 0.2 and outperforms MA and Weighted MA. In addition, MA and Weighted MA perform worse when $n_\text{batch}$ becomes small (i.e., fewer microtasks in a batch) or when $k$ is large, which means that the distance between two consecutive batches does not represent the distance between the current state and the final state. 

% No matter in which situation, $\Delta_{\text{SC}}$ of AnsBoost is less than or equal to $0.2$ and smaller than $\Delta_{\text{SC}}$ of MA and Weighted MA. Especially, MA and Weighted MA are ineffective when $k = 20$ because $t_{\text{acc}}$ is too loose (the number of maximal swaps is 8 in top-$20$ list). It is easy to make the distance among different batches smaller than $t_\text{acc}$ and leads that MA and weighted MA stop very early. Besides, MA and Weighted MA perform badly when $n_\text{batch} = 50$ and $100$ because the distance among different batches is small which is hard to represent the distance between the current result and the final result. It also leads that MA and Weighted MA stop early. 

\begin{figure}[ht!]
\begin{minipage}[t]{0.5\textwidth}
 \centering
   \includegraphics[width=1\textwidth]{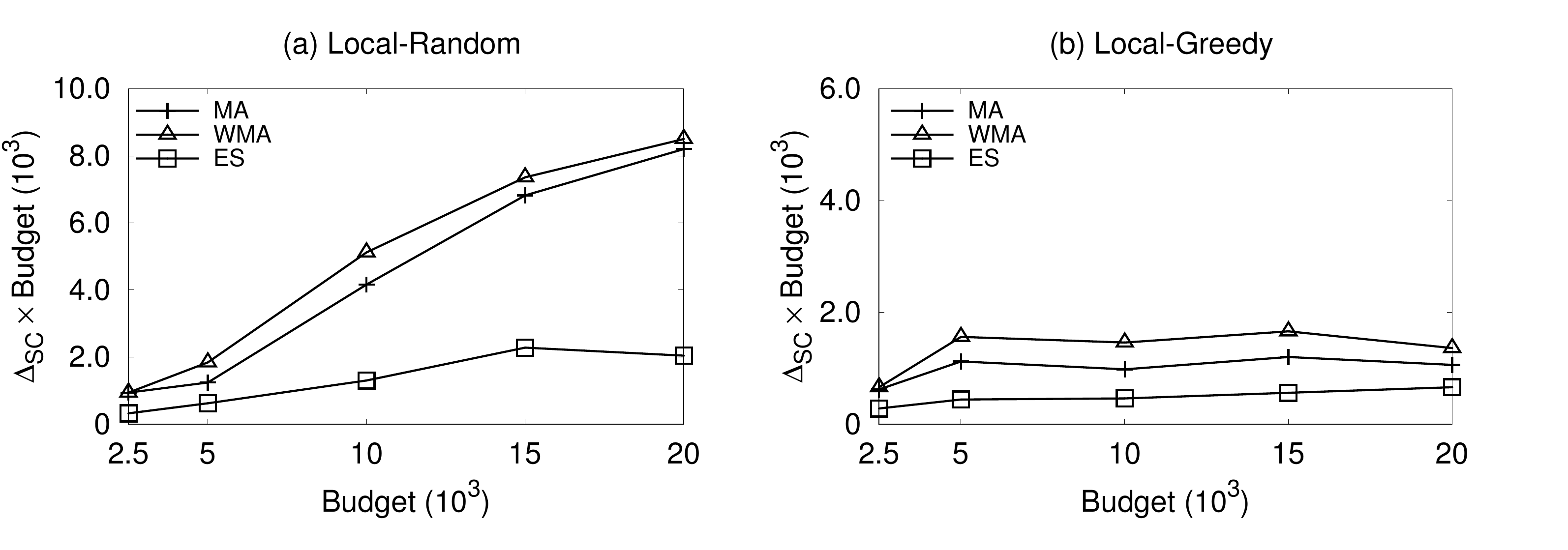}
  \caption{$\Delta_{\text{SC}}$ in varied $B$}
  \label{fig:diff_budget} 
\end{minipage}
 % \centering
 %   \includegraphics[width=0.47\textwidth]{figs/diff_budget_absolute.eps}
 %  \caption{$\Delta_{\text{SC}}$ in varied $B$}
 %  \label{fig:diff_budget} 
\end{figure}

Fig.~\ref{fig:diff_budget} evaluates the effect of the budget $B$. Note that we use the absolute number of answers instead of a percentage in the y-axis. We set $\theta=0.02$ and $n_\text{batch}=200$ as default. We try $2.5*10^3$, $5.0*10^3$, $10.0*10^3$ and $20.0*10^3$ for the budget. $\Delta_\text{sc}$ of ES is always smaller than the corresponding $\Delta_\text{sc}$ of MA and Weighted MA. Particularly, errors of MA and Weighted MA increase dramatically when $B$ increases in {\it Local-Random}. This is because increasing budget B improves the quality of the final result and the position of the optimal stopping point moves backwards. But the stopping points predicted by MA and Weighted MA do not change.

%In {\it Greedy}, used budget by the BSP is not changed with increasing budget because the final result is not changed. It means that {\it Local-Greedy} already knows enough information for top-$k$ items when $|L|$ is $5.0*10^3$. The misused budget by our method AnsBoost is always smaller compared to MA and weighted MA.

\begin{table}[t!]

%begin{minipage}{0.48\linewidth}
\centering
\caption{Avg. Used Budget (UB) and Actual Error (AE) for Top-$k$}
\label{table:avg_saved_budget_topk_list}
\begin{tabular}{cccccc}
\toprule
\multicolumn{2}{c}{\multirow{2}{*}{Top-k List}} & \multicolumn{2}{c}{PeopleNum Dataset} & \multicolumn{2}{c}{PeopleAge Dataset} \\ \cline{3-6} 
\multicolumn{2}{c}{}                            &  UB  &  AE  &  UB  &  AE  \\
\midrule
\multirow{3}{*}{$\theta = 0.005$}   & Optimal         & 88\%              & 0.000             & 94\%              & 0.000             \\ \cline{2-6} 
                                          & ES   & 87\%              & 0.003             & 95\%              & 0.002             \\ \cline{2-6} 
                                          & WMA         & 54\%              & 0.007             & 96\%              & 0.004             \\ \hline
\midrule
\multirow{3}{*}{$\theta = 0.01$}    & Optimal         & 77\%              & 0.008             & 88\%              & 0.006             \\ \cline{2-6} 
                                          & ES   & 78\%              & 0.009             & 92\%              & 0.005             \\ \cline{2-6} 
                                          & WMA         & 57\%              & 0.027             & 84\%              & 0.016             \\ \hline
\midrule
\multirow{3}{*}{$\theta = 0.02$}    & Optimal         & 66\%              & 0.018             & 82\%              & 0.015             \\ \cline{2-6} 
                                          & ES   & 58\%              & 0.028             & 82\%              & 0.019             \\ \cline{2-6} 
                                          & WMA         & 26\%              & 0.052             & 49\%              & 0.068             \\ \hline
\midrule
\multirow{3}{*}{$\theta = 0.03$}    & Optimal         & 54\%              & 0.024             & 77\%              & 0.024             \\ \cline{2-6} 
                                          & ES   & 45\%              & 0.038             & 74\%              & 0.031             \\ \cline{2-6} 
                                          & WMA         & 20\%              & 0.055             & 36\%              & 0.089             \\ \hline
\bottomrule
\end{tabular}
\end{table}

%\end{minipage}

\begin{table}[t!]
%\begin{minipage}{0.48\linewidth}  
\centering

%\begin{table}[]\scriptsize
\caption{Avg. Used Budget and Actual Error for Ranking}
\label{table:avg_saved_budget_ranking}
\begin{tabular}{cccccc}
\toprule
\multicolumn{2}{c}{\multirow{2}{*}{Ranking}} & \multicolumn{2}{c}{PeopleNum Dataset} & \multicolumn{2}{c}{PeopleAge Dataset} \\ \cline{3-6} 
\multicolumn{2}{c}{}                            &  UB  &  AE  &  UB  &  AE  \\
\midrule
\multirow{3}{*}{$\theta= 0.005$}   & Optimal         & 95\%              & 0.002             & 95\%              & 0.003             \\ \cline{2-6} 
                                          & ES   & 92\%              & 0.004             & 96\%              & 0.003             \\ \cline{2-6} 
                                          & WMA         & 91\%              & 0.006             & 96\%              & 0.001             \\ \hline
\midrule
\multirow{3}{*}{$\theta = 0.01$}    & Optimal         & 89\%              & 0.006             & 90\%              & 0.007             \\ \cline{2-6} 
                                          & ES   & 82\%              & 0.015             & 88\%              & 0.010             \\ \cline{2-6} 
                                          & WMA         & 60\%              & 0.035             & 61\%              & 0.029             \\ \hline
\midrule
\multirow{3}{*}{$\theta = 0.015$}   & Optimal         & 83\%              & 0.013             & 85\%              & 0.011             \\ \cline{2-6} 
                                          & ES   & 70\%              & 0.027             & 81\%              & 0.019             \\ \cline{2-6} 
                                          & WMA         & 29\%              & 0.065             & 35\%              & 0.051             \\ \hline
\midrule
\multirow{3}{*}{$\theta = 0.02$}    & Optimal         & 79\%              & 0.016             & 80\%              & 0.017             \\ \cline{2-6} 
                                          & ES   & 61\%              & 0.033             & 72\%              & 0.028             \\ \cline{2-6} 
                                          & WMA         & 19\%              & 0.067             & 31\%              & 0.057             \\ \hline
\bottomrule
\end{tabular}
%\end{minipage}
\end{table}

\begin{figure*}[ht!]
   \includegraphics[width=1\textwidth]{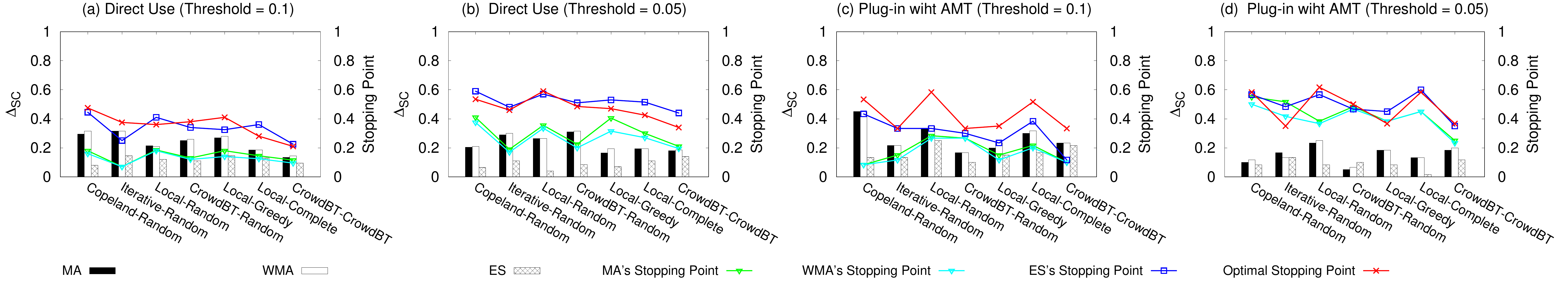}
  \caption{$\Delta_{\text{sc}}$ \& Stopping Points in Online Experiments}
  \label{fig:demo_all_fig} 
\end{figure*}

%!TEX root = main.tex

\section{Online Experiments In Prototype Systems}\label{sec:case_study}

To show the effectiveness in practice, we designed an online crowdsourcing system that collects microtask answers from real workers.

\subsection{Prototype System} \label{sec:prototype}

\begin{figure}[tb!]
  \begin{subfigure}[b]{0.98\columnwidth}
    \centering
    \includegraphics[height=2.54in]{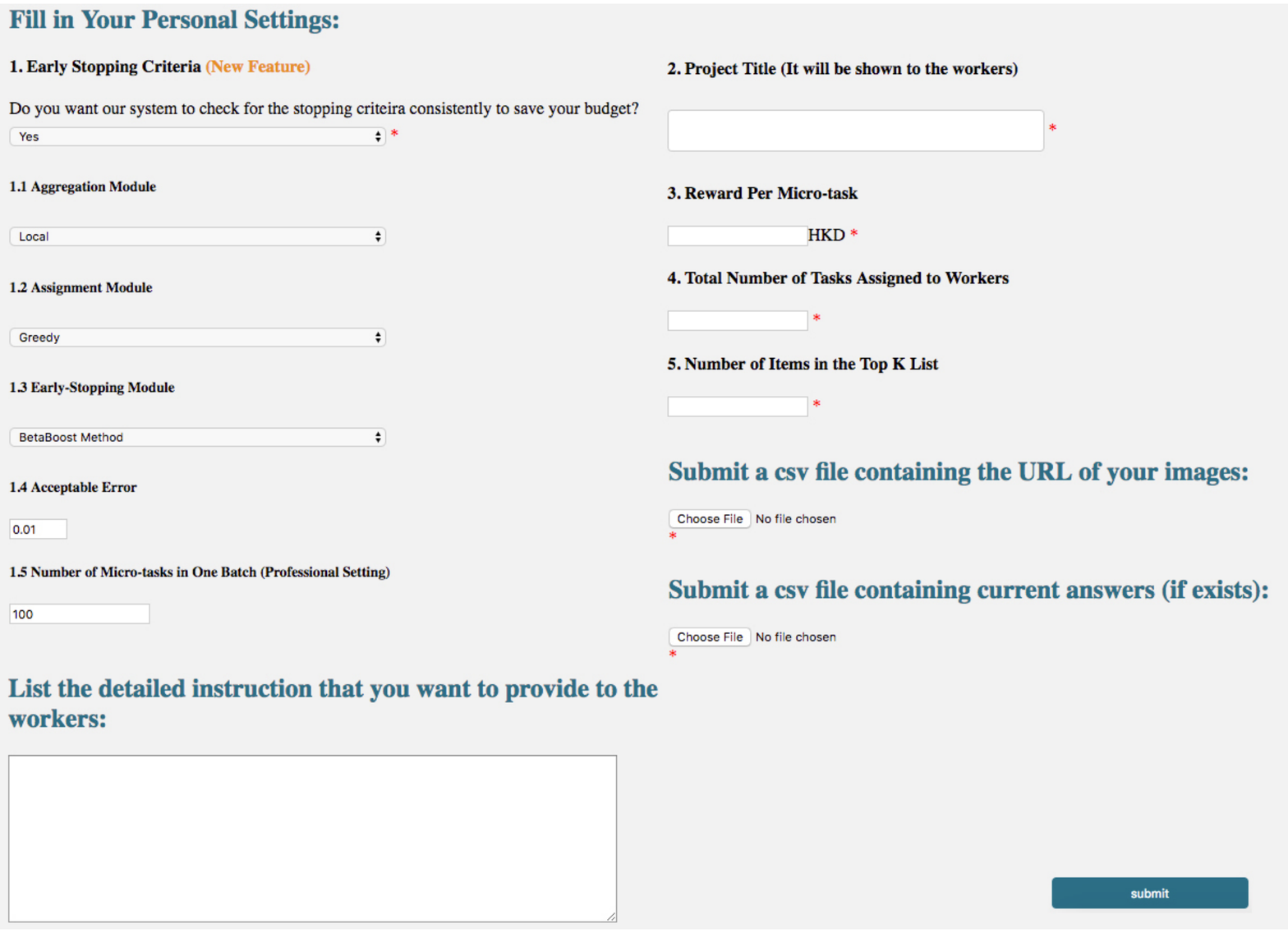}
    \caption{Interface}
    \label{fig:framework_demo_a}
  \end{subfigure}\\
  \begin{subfigure}[b]{0.485\columnwidth}
    \centering
    \includegraphics[width=1.48in]{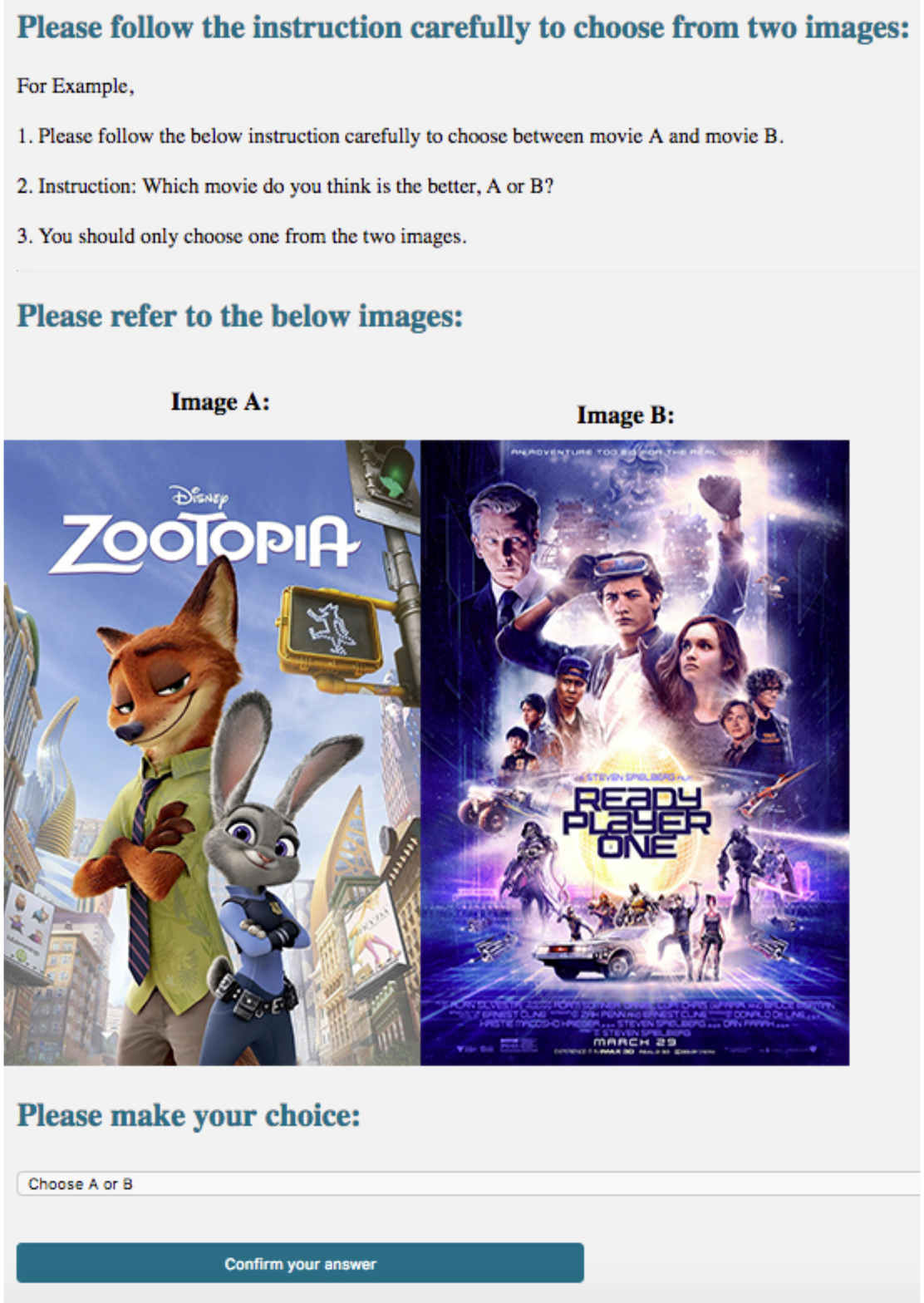}
    \caption{Interface}
    \label{fig:framework_demo_b}
  \end{subfigure}%
  \begin{subfigure}[b]{0.485\columnwidth}
    \centering
    \includegraphics[width=1.85in]{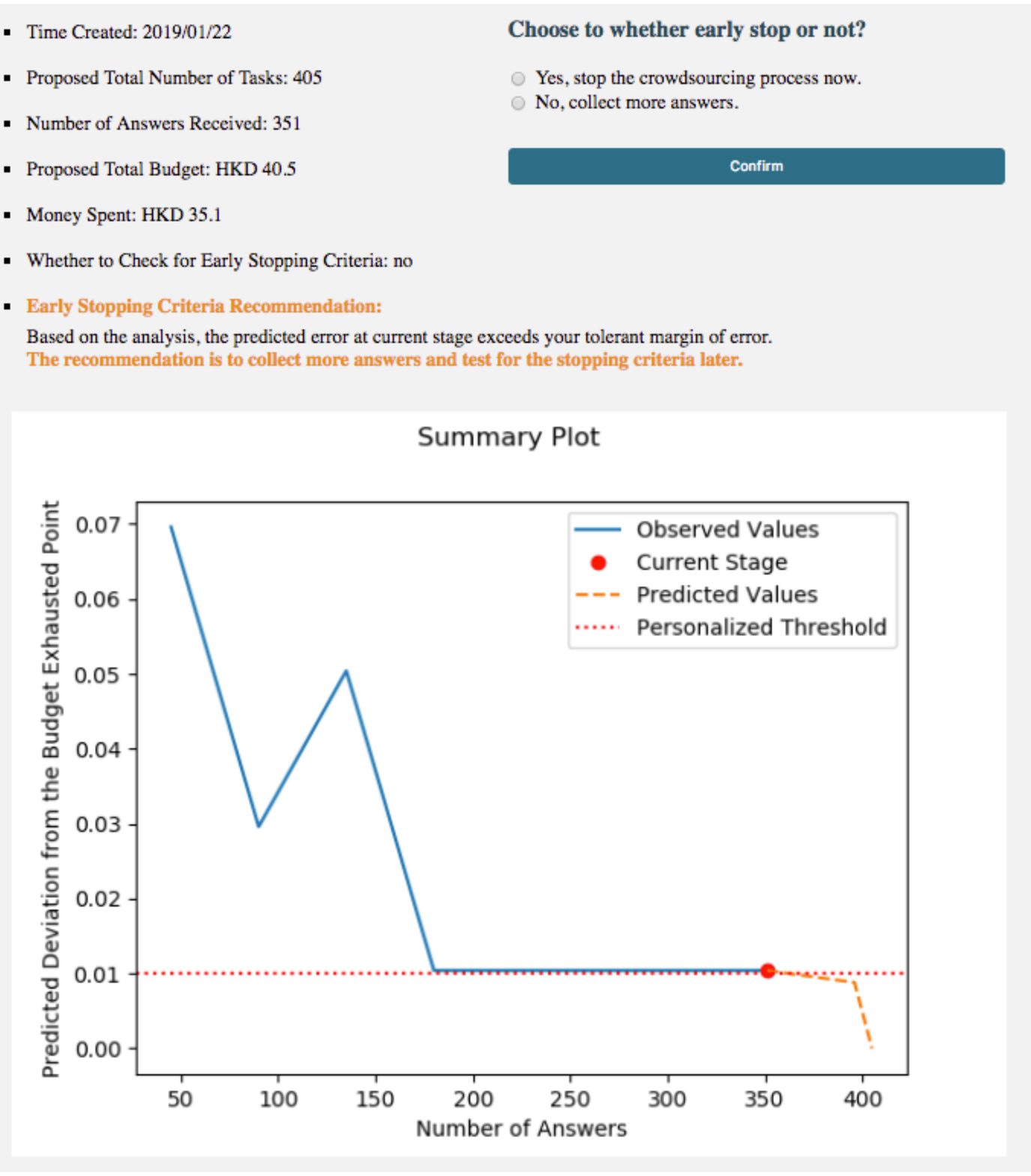}
    \caption{Interface}
    \label{fig:framework_demo_c}
  \end{subfigure}%
  \caption{Prototype System}
\end{figure}

%%%%%%%%%%%%%%%%%%%%%%%%%%%%%%%%%%%%%%%%%%%%%%%%%%%%%%%
%%%%%%%  add the figures of interfaces   %%%%%%%%%%%%%%%%%%
%%%%%%%%%%%%%%%%%%%%%%%%%%%%%%%%%%%%%%%%%%%%%%%%%%%%%%%
Fig.~\ref{fig:framework_demo_a} - \ref{fig:framework_demo_c} show the interface of our prototype, which follows the website style of AMT.
Requesters create a new task using the form shown in Fig.~\ref{fig:framework_demo_a}. %\footnote{http://chshan-server.cs.hku.hk/demo/create-tasks.php}
They can write down the title and descriptions of tasks, upload the objects, choose specific aggregation, assignment and early-stopping modules and set the corresponding parameters.
We provide four default aggregation methods ({\it Copeland}, {\it Iterative}, {\it Local} and {\it Crowdbt}) and four default assignment methods ({\it Random}, {\it Greedy}, {\it Complete} and {\it Crowdbt}).
%Requesters may also use their own aggregation and assignment methods via the provided C++ interface.
%Detailed instructions are listed on the web page.
After setting up the problem, requesters
%could
obtain IDs for the tasks and corresponding microtask links;
see for example Fig.~\ref{fig:framework_demo_b}. %
%\footnote{http://chshan-server.cs.hku.hk/demo/answer-submit.php?projectID=0}
%a ID for the task. For example, micro-tasks are generated and could be accessed on the page (b) \footnote{http://chshan-server.cs.hku.hk/demo/answer-submit.php?projectID=9} if the ID is 9.
Requesters can either post the microtask links on a commercial platform (e.g., AMT) or simply distribute the microtasks to volunteers via a local platform.

After collecting a part of answers, requesters can download the current answer set, obtain the inferred results, and see the curve for the predicted expected distance between the current result and the final result, as shown in Fig.~\ref{fig:framework_demo_c}. %\footnote{http://chshan-server.cs.hku.hk/demo/manage.php?projectID=0} is an example web page.
The plot is drawn by Python and each point is the expected distance between the inferred ranking after each batch and the final result when the budget is exhausted.
We use solid (dotted) line before (after) the current point.

Besides creating a new task and asking workers to answer, our system also provides an interface that allows to upload collected answer set and analyze the performance by tuning different parameters in Fig.~\ref{fig:framework_demo_a}.
Similarly, users can see the inferred results and their expected distance in the form of Fig.~\ref{fig:framework_demo_c}.

\subsection{Local System Experiments} \label{sec:direct_use}

In our local system, we attempt to rank top-5 movies among 10 popular ones, such as Zootopia and Ready Player.
Each microtask asks a worker to select the better one between two movies.
To run the inference model, we set the default budget to 300 and the number of microtasks in a batch to 20. Our experiments invite 18 volunteers to participate.
Our system dispatches microtasks and infers ranking by seven assignment and inference models {\it Copeland-Random}, {\it Iterative-Random}, {\it Local-Random}, {\it CrowdBT-Random}, {\it Local-Greedy}, {\it Local-Complete}, and {\it CrowdBT-CrowdBT}. As different worker sequences and reliability distributions may affect the performance, we ran each assignment method 5 times so that we obtained 35 answer curves in total.

%For the answer sequences collected by {\it Random}, we can infer the ranking by four algorithms, e.g., {\it Copeland}, {\it Iterative}, {\it Local} and {\it Crowdbt}. For the others, we can only infer the ranking by its specific algorithm, e.g., the sequence collected by {\it Local} can only be inferred by {\it Greedy}.

%The default budget is 300 and the number of micro-tasks in a batch is 20. As different worker sequences and reliability distributions may affect the performance of our strategies, we ran each assignment method several times.

%Our experiments invite 18 volunteers to answer the assigned micro-tasks. We tried {\it Copeland}, {\it Iterative}, {\it Local} and {\it Crowdbt} for the worker answer sequences decided by the {\it Random} assignment method. At the end, we collect 35 answer sequences in total.
%We ran the four assignment methods 5 times and obtained 20 answer sequences. Though applying four inference modules into {\it Random}, we totally have 35 distance curves.

We calculated the optimal stopping point $p_\text{optimal}$ and the stopping point $p_\text{sc}$ decided by MA, Weighed MA and ES for these distance curves. We set the window size of MA and Weighed MA to $w=5$ and $\theta$ to 0.1 and 0.05.
Fig.~\ref{fig:demo_all_fig}(a) and (b) show $\Delta_\text{sc}$ and the stopping point in different situations. ES's $\Delta_\text{sc}$ is always smaller than those of MA and Weighted MA. Moreover, ES's stopping point is also always closer to the optimal stopping point.

\subsection{Amazon Mturk Experiments}

We also ran the same experiments on Amazon Mechanical Turk (AMT). We attach the link of our local system in the microtasks and we pay $\$0.01$ for 5 pairwise comparisons. A worker is qualified to answer the microtasks only if her HIT approval rate is greater than $90\%$. We ran the seven assignment and inference models 3 times and obtained 21 distance curves.

Similar to the experiment in Sec.~\ref{sec:direct_use}, we test different stopping criteria when the window size $w$ is $5$ and the accuracy tolerance $\theta$ is $0.1$ or $0.05$. Fig.~\ref{fig:demo_all_fig}(c) and (d) show $\Delta_\text{sc}$ and the stopping point in different situations. ES's $\Delta_\text{sc}$ is smaller than those of MA and Weighted MA in most cases except for {\it Crowdbt-Random} when $\theta= 0.05$. Compared with MA and Weighted MA, ES is closer to the optimal stopping point on average.

%!TEX root = main.tex

\section{related work}\label{sec:relatedwork}
%  The related work is divided into two categories: crowdsourced ranking and  early-stopping strategies in crowdsourcing.
%area and general settings
In this section, we review the related work from two categories: crowdsourced ranking  and early-stopping strategies for crowdsourcing.

\stitle{Crowdsourced ranking} 
The ranking problem has a long history and has been studied in the past several decades. Simple traditional ranking algorithms, e.g., BordaCount \cite{BordaCount} and Copeland \cite{Copeland}, rank objects by the times they win/lose in the comparisons. In the context of crowdsourcing, we also need to consider the fact that the crowd may give incorrect answers (traditional ranking algorithms do not consider possible errors). How to deal with noisy answers and control worker qualities is the key component in almost all crowdsourcing problems \cite{surveycrowdsourced, surveytruthinference, shantabularshort, shan2019tabular}. To solve it, {\it inference algorithms} have been proposed to infer a ranked list based on the collected (noisy) answers. Moreover, {\it task assignment strategies}  have been put forward to allocate suitable tasks to high-quality workers and obtain informative answers. 

Inference algorithms in raking problems can be divided into two categories: heuristic-based solutions from the DB community approach the problem as a top-$k$ operation in databases, and machine-learning algorithms formalize it as a leaning problem and maximize the likelihood of top-$k$ objects. Heuristic score-based algorithms \cite{local,iterative,BRE,venetis2012max} rank objects by estimating the underlying score of objects. CrowdBT~\cite{crowdbt} and CrowdGauss~\cite{crowdgauss} are ML algorithms, which set the objective function based on the assumption and use maximum likelihood to obtain the top-$k$ object with the highest probability.  

%in order to
%compared objects and
%collect informative answers
% under the given budget (task selection).
% For answer aggregation, simple election algorithms, like BordaCount \cite{BordaCount} and Copeland \cite{Copeland}, can be used.
% Score-based algorithms \cite{local} determine a score for each object and
% then rank them based on it.
%the top-$k$ result is k objects with the highest scores. 
%There are also approaches \cite{local} which iteratively eliminate the low rank objects until $k$ objects left. 
% There are also machine learning methods, such as CrowdBT \cite{crowdbt} and CrowdGauss \cite{crowdgauss}, which use maximum likelihood to estimate the top-$k$ object set that has the highest probability.  
Regarding task assignment strategies, Amazon MTurk follows a random assignment strategy;
i.e., microtasks are randomly dispatched to each coming worker.
%and guarantee each microtask has been answered the same number of times. 
%In the top-$k$ situation, requesters should firstly decide the format of the question, pairwise comparison or multi-wise comparison. If a question has $m$ objects waiting for workers to compare, there are $C_n^m$ questions and Amazon Mturk strategy will make them be answered the same number of times. 
Random assignment does not consider the difficulties of microtasks.
Some heuristic assignment methods \cite{local} aim at maximizing the probability of obtaining the top-$k$ result, e.g., by selecting most promising object pairs (e.g., with the largest latent scores) to compare.
%or making all the objects be compared.
\cite{ssrw} avoids some unnecessary comparisons by setting a bound for the object latent score. Active learning methods are also used in CrowdBT~\cite{crowdbt} and CrowdGauss~\cite{crowdgauss} to compare objects with the largest expected information gain. 

According to a recent experimental study \cite{exptopk},
different inference and assignment
methods have their own advantages
and there does not exist a single best one.
%there is no exist the best aggregation and assignment method for all scenarios. Different methods have their own advantages and disadvantages.
Machine-learning methods for answer aggregation typically have high quality.
%For answer aggregation, the machine-learning methods achieve high quality.
Still, global inference heuristics that utilize global comparison results achieve comparable and even higher quality than ML methods.
Local inference heuristics have poor quality, but have higher efficiency and scalability. For task assignment, active-learning methods achieve higher quality than heuristics, but they have low efficiency.
%and cannot meet the online requirement. 

\stitle{Stopping Criteria}
Stopping criteria have been defined for various crowdsourcing problems. \cite{mo2013optimizing} and \cite{raykar2014sequential} design an early-stopping strategy
%with the help of a utility function and a pre-set threshold
for
%the
multiple-choice-question problems (e.g., choosing the opinion positive, neutral, or negative in a sentence). \cite{welinder2010online} and \cite{li2017optimal} use Sequential Probability Ratio Test (SPRT) \cite{wald1945sequential} to decide when to stop for multi-labeling tasks (e.g., labeling pictures as a portrait or a landscape). Besides, \cite{chai2018incentive} uses Chao92 \cite{chao1992estimating} estimator to estimate the level of completion (and the termination point) for entity collection problems (e.g., collecting a set of active NBA players). The settings of all these problems are quite different from crowdsourced ranking. The reason is that microtasks are independent in these problems, while microtasks in ranking problems are correlated.
%We cannot ignore the correlation among microtasks which is important for us to infer the order. 

Previous work on crowdsourced ranking \cite{kou2017crowdsourced} and \cite{chen2017asymptotically} define stopping conditions in their assignment methods.
%for crowdsourced rank or top-$k$ list problem..
For instance, \cite{chen2017asymptotically} assumes that each object has a latent score and answers to pairwise comparisons follow the Bradley-Terry (BT) \cite{BTL} model.
However, this approach cannot be used with methods whose assumption is the Thurstone's model \cite{thurstoneModel} or Plackett-Luce model \cite{PlackettLuce}. It is also not suitable if the objects do not have scores.
%for inference models where there are no scores for objects. 
\cite{kou2017crowdsourced} asks the crowd to give a numerical answer in $[0,1]$ for a pairwise comparison and calculates the confidence interval of the result. This approach cannot be applied if the answers are just 0 or 1, and it is based on special assumptions that cannot generalize to most situations.

%A straightforward idea is to measure the change between two consecutive batches and stop if the change is smaller than a threshold. However, the result may be oscillatory convergent, i.e., it may be stable during a small period of time and then change as the number of answers increases. In this case, we may stop earlier than desired.
Moving average and Weighted moving average \cite{MovingAverage} are to measure the change between two consecutive batches and stop if the change is smaller than a threshold. However, the result may be oscillatory convergent, i.e., it may be stable during a small period of time and then change as the number of answers increases. In this case, they stop earlier than desired. Besides, it is hard for users to set the best parameter values for them, such as the window size. Bad parameters lead to the worst stopping position.

%!TEX root = main.tex

\section{conclusion}\label{sec:conclusion}
In this paper, we proposed a general stopping criterion for crowdsourced ranking. The goal is to terminate answer collection as soon as the estimated ranking becomes very similar to the ranking that we expect to obtain when all the budget is exhausted. This way, resources and time for obtaining the result are saved.
%to calculate the difference between the current and the final result for crowdsourced rank or top-$k$ list problem.
We demonstrated the robustness of our method in different situations, including subjective or objective tasks, diverse inference modules or task assignment modules and different budget and tolerance parameter values. We implemented a prototype which can be used by requesters in practice.

There are several directions for future work. First of all, although we have only considered pairwise comparisons as microtasks in this paper, our module can also be applied for microtasks where multiple objects are compared by a worker; in this case, we use multinomial distributions instead of Bernoulli distributions. Besides, some inference modules, such as rating-based algorithms\cite{rating1} and hybrid algorithms\cite{khan2014hybrid}, are still not suitable for our current module. We will study the extension of our module to predict ratings of objects.

% \author{Caihua Shan}
% \affiliation{
%  \institution{The University of Hong Kong}
%  }
% \email{chshan@cs.hku.hk}

% \author{Leong Hou U}
% \affiliation{
%  \institution{University of Macau}
%  }
% \email{ryanlhu@umac.mo}

% \author{Nikos Mamoulis}
% \affiliation{
%  \institution{University of Ioannina}
%  }
% \email{nikos@cs.uoi.gr}

% \author{Reynold Cheng}
% \affiliation{
%  \institution{The University of Hong Kong}
%  }
% \email{ckcheng@cs.hku.hk}

% \author{Ruohan Li}
% \affiliation{
%  \institution{The University of Hong Kong}
%  }
% \email{u3523358@connect.hku.hk}

% \pagestyle{plain}
% \pagenumbering{arabic}

%\author{ Caihua Shan{$\,^\dag$}~~~Leong Hou U{$\,^\#$}~~~Nikos Mamoulis{$\,^\sharp$}~~~Reynold Cheng{$\,^\dag$}}

% \vspace{1.6mm}\\
% {$\,^\dag$}\, Department of Computer Science, The University of Hong Kong \\
% {$\,^\#$}\, Department of Computer Science, University of Macau\\
% {$\,^\sharp$}\, Department of Computer Science, University of Ioannina \\

% \{chshan, ckcheng\}@cs.hku.hk,~~~ryanlhu@umac.mo,~~~nikos@cs.uoi.gr

% \section*{Acknowledgements}
% We would like to thank the reviewers for the insightful comments. 
% We also want to thank RuoHan Li for helping us build the prototype system.
% Xiang Li is the corresponding author of this paper. 
% Reynold Cheng and Caihua Shan were supported by the Research Grants Council of Hong Kong (RGC Projects HKU 17229116 and 17205115) and the University of Hong Kong (Projects 104004572, 102009508, 104004129). 

% \balance
{ 
\bibliographystyle{abbrv} 
\bibliography{refs/paper}
}

\end{document}